\documentclass[12pt,a4paper]{article}

\usepackage[skins,theorems]{tcolorbox}
\tcbset{highlight math style={enhanced,
  colframe=red,colback=white,arc=0pt,boxrule=1pt}}
  \usepackage[bookmarksopen, bookmarksnumbered, bookmarksopenlevel=2]{hyperref}
  \usepackage{tikz}
  \usepackage{tikz-3dplot}
 \usetikzlibrary{calc}
 \usetikzlibrary{decorations} %
 \usepackage[UKenglish]{babel}
 \usepackage[toc,page]{appendix}
 \usepackage{amsmath}
 \usepackage{amssymb}
 \usepackage{graphicx}
 \usepackage{hhline}
 \usepackage[bf]{caption}
\usepackage{cite}
\usepackage[vcentermath]{youngtab}
\usepackage{geometry}
\usepackage{slashed}
\usepackage{color}
\usepackage{stackrel}
\usepackage{tikz}
\usepackage{tikz-cd} 
\usepackage{empheq}

\newcommand{\bqa}{\begin{eqnarray}}
\newcommand{\eqa}{\end{eqnarray}}

\newcommand{\nn}{\nonumber}

\def\otherK3{{{\cal K}}   }
\def\Kcone{{\bf K}}
\def\Mcone{{\bf M}}

\hypersetup{
    pdftitle={},
    pdfauthor={},
    pdfsubject={}
}
\numberwithin{equation}{section}
\numberwithin{table}{section}

\makeatletter

\newcommand{\be}{\begin{equation}}
\newcommand{\ee}{\end{equation}}
\newcommand{\beq}{\begin{equation}}
\newcommand{\eeq}{\end{equation}}
\newcommand{\ba}{\begin{aligned}}
\newcommand{\ea}{\end{aligned}}

\newcommand{\bea}{\begin{eqnarray}}
\newcommand{\eea}{\end{eqnarray}}

\newcommand{\cO}{\mathcal{O}}

\newcommand{\cE}{\mathcal{E}}

\newcommand{\cN}{\mathcal{N}}

\newcommand{\cF}{\mathcal{F}}
\newcommand{\cI}{\mathcal{I}}

\def\unit{{1\kern-.65ex {\rm l}}}
\def\1{{1\kern-.65ex {\rm l}}}

\def\IZ{\mathbb{Z}}
\def\IP{\mathbb{P}}

\def\IR{{\mathbb{R}}}

\newcount\hour \newcount\minute
\hour=\time \divide \hour by 60
\minute=\time
\def\now{%
\ifnum \hour<13
  \ifnum \hour=0 \advance \hour by 12 \number\hour:\else \number\hour:\fi%
     \ifnum \minute<10 0\fi%
     \number\minute%
\ A.M.%
\else \advance \hour by -12 \number\hour:%
  \ifnum \minute<10 0\fi%
  \number\minute%
  \ P.M.%
\fi%
}

\def\fnote#1#2{\begingroup\def\thefootnote{#1}\footnote{#2}
     \addtocounter{footnote}{-1}\endgroup}

\makeatother

\begin{document}






\begin{flushright}
{\tt\normalsize CERN-TH-2018-220}\\
\end{flushright}

\vskip 40 pt
\begin{center}
{\large \bf A Stringy Test of the Scalar Weak Gravity Conjecture} 

\vskip 11 mm

Seung-Joo Lee${}^{1}$, Wolfgang Lerche${}^{1}$,
and Timo Weigand${}^{1,2}$

\vskip 11 mm

\small ${}^{1}${\it CERN, Theory Department, \\ 1 Remoulade des Particules, Geneva 23, CH-1211, Switzerland} \\[3 mm]
\small ${}^{2}${\it Institut f\"ur Theoretische Physik, Ruprecht-Karls-Universit\"at, \\
Philosophenweg 19, 69120 Heidelberg, Germany}

\fnote{}{seung.joo.lee, wolfgang.lerche,  
timo.weigand @cern.ch}

\end{center}

\vskip 7mm

\begin{abstract}

We prove a version of the Weak Gravity Conjecture for 6d F-theory or heterotic string compactifications with 8 supercharges.
This sharpens our previous analysis by including massless scalar fields.  
The latter are known to modify  the Weak Gravity Conjecture bound in two a priori independent ways:
First, the extremality condition of a charged black hole is modified, and second,
the test particles required to satisfy the Weak Gravity Conjecture are subject to additional Yukawa type interactions.
We argue on general grounds that at weak coupling, the two types of effects are 
equivalent for a  tower of asymptotically massless charged test particles predicted by the Swampland Distance Conjecture.

We then specialise  to F-theory compactified on elliptic Calabi-Yau three-folds  
and prove that the precise numerical bound on the charge-to-mass ratio is satisfied at weak coupling.
 This amounts to an intriguing coincidence of two a priori different notions of extremality, namely one based on the balance of gauge, gravitational and scalar forces for extremal (non-BPS) black holes, and the other encoded in the modular properties of certain Jacobi forms. 
In the presence of multiple abelian gauge group factors, the elliptic genus counting these states is a lattice quasi-Jacobi form of higher rank, and we exemplify this  in a model with two abelian gauge group factors.

\end{abstract}

\vfill

\thispagestyle{empty}
\setcounter{page}{0}
\newpage

\tableofcontents



\section{\large Introduction and Summary}

In a consistent quantum theory,  the gravitational and the non-gravitational sectors are typically far from independent.
String theory as a framework for a UV consistent theory is the prime example where this principle is realized.
More generally, a  number of remarkable conjectures constrain the properties of quantum field theories by requiring that they can be consistently coupled to quantum gravity.
One of the earliest such conjectures is the Weak Gravity Conjecture (WGC) \cite{ArkaniHamed:2006dz} and its refinements \cite{Cheung:2014vva,Heidenreich:2015nta,Heidenreich:2016aqi}. It postulates the existence of a set of particles on which gravity acts weaker than any gauge force.
Apart from being an exciting arena to study general principles of fundamental physics, the conjecture has important implications on phenomenology. 
At a fundamental level, it is related to the so-called Swampland Distance Conjecture \cite{Ooguri:2006in}, as studied in \cite{Klaewer:2016kiy,Palti:2017elp,Heidenreich:2017sim,Andriolo:2018lvp,Heidenreich:2018kpg,Grimm:2018ohb,Lee:2018urn}, guarantees Weak Cosmic Censorship in AdS space \cite{Crisford:2017gsb,Cottrell:2016bty} and underlies a conjectured instability of non-supersymmetric AdS vacua \cite{Ooguri:2016pdq,Freivogel:2016qwc,Danielsson:2016mtx}.
From a more applied point of view, its generalization to $p$-form interactions for $p = 0$ 
constrains theories of large field inflation, as reflected in a substantial corpus of work including \cite{Rudelius:2014wla,delaFuente:2014aca,Rudelius:2015xta,Montero:2015ofa,Brown:2015iha,Bachlechner:2015qja,Hebecker:2015rya,Brown:2015lia,Heidenreich:2015wga,Kooner:2015rza,Ibanez:2015fcv,Hebecker:2015zss,Baume:2016psm,Heidenreich:2016jrl,Blumenhagen:2017cxt,Valenzuela:2017bvg,Ibanez:2017vfl,Aldazabal:2018nsj,Blumenhagen:2018nts,Blumenhagen:2018hsh,Reece:2018zvv} (for additional aspects of this axionic Weak Gravity Conjecture see \cite{Hebecker:2017uix,Montero:2017yja,Hebecker:2018ofv}); other versions affect the allowed range of St\"uckelberg masses \cite{Reece:2018zvv} and have been argued to have implications on neutrino physics \cite{Ooguri:2016pdq,Ibanez:2017kvh,Hamada:2017yji} or the electroweak scale \cite{Ibanez:2017oqr,Lust:2017wrl,Gonzalo:2018dxi}. 
Understanding the WGC and its refinements in as much detail as possible is therefore an important subject.  General arguments aiming to prove the Weak Gravity Conjecture involve 
the entropy of black holes \cite{Cottrell:2016bty,Fisher:2017dbc,Cheung:2018cwt}, consistency of the effective field  theory \cite{Andriolo:2018lvp,Hamada:2018dde} or the AdS/CFT correspondence \cite{Nakayama:2015hga,Harlow:2015lma,Montero:2016tif,Benjamin:2016fhe}.
A lot of evidence for the various forms of the conjecture has already been accumulated in a string theoretic context e.g. in \cite{ArkaniHamed:2006dz,Heidenreich:2016aqi,Grimm:2018ohb}, in situations with extended supersymmetry in various numbers of dimensions.

\subsection{Recap of previous analysis}

In \cite{Lee:2018urn} we have studied the Weak Gravity Conjecture for 6d string compactifications with {\it minimal}, i.e. with $N=(1,0)$, supersymmetry. 
As noted above, the WGC is related to the so-called Swampland Distance Conjecture \cite{Ooguri:2006in}, which posits that as we traverse the moduli space of a theory over infinite distances, a tower of states must become exponentially light.
One such  regime is where the gauge coupling becomes asymptotically weak. Near this region, infinitely many charged massless states are expected to appear \cite{Heidenreich:2017sim,Andriolo:2018lvp,Heidenreich:2018kpg,Grimm:2018ohb} and to lead to a breakdown of the effective theory. This is consistent with the postulate \cite{Banks:1988yz,Banks:2010zn,Montero:2017mdq} that no global symmetries can arise in a theory with dynamical gravity.

In \cite{Lee:2018urn} we have characterized the most general limit in 6d $N=(1,0)$ supersymmetric compactifications of F-theory in which the gauge coupling of a single gauge group factor becomes asymptotically weak, while gravity stays dynamical. 
In this limit, a curve $C$ on the base $B_2$ of the elliptic fibration $Y_3$ tends to infinite volume while the volume of $B_2$ stays finite.
The expected tower of charged asymptotically massless states has been explicitly identified: It consists of  certain {\it non-BPS} excitations of a D3-brane wrapped on a  curve of zero self-intersection. The volume of the wrapped curve tends to zero in the weak coupling limit, and the wrapped D3-brane gives rise to a string which becomes tensionless as the gauge coupling vanishes. This asymptotically tensionless string is in fact the 6d critical heterotic string. Its elliptic genus \cite{Schellekens:1986yi,Witten:1986bf} captures a subset of the physical string excitation spectrum and can be computed, by duality with M-theory \cite{Klemm:1996hh,Haghighat:2013gba,Haghighat:2014vxa}, in terms of the genus-zero Gromov-Witten invariants of the elliptic three-fold,  $Y_3$. The key observation  \cite{Lee:2018urn} is that, for a theory with a single $U(1)$ factor, the  spectrum includes states whose charges span the sublattice 
\be \label{sublattice1}
\mathfrak{q}_k = 2m \, k\,\,, \qquad k \in \mathbb Z \,,
\ee
of the full charge lattice, and whose masses $M_k$ satisfy the relation\footnote{A priori, this relation holds in the weak gauge coupling regime, where the dual heterotic string becomes perturbative as well.}
\be \label{foundrel1}
\mathfrak{q}_k^2 =  4 m \, n_k = 4m \, (M_k^2/ (4m g_{\rm YM}^2) + 1) \,.
\ee
Here $m$ is an integer of order one which can be read off from the geometric realisation of the theory, and which coincides with the fugacity index of the elliptic genus. Moreover, $n_k$ denotes the string excitation level at which the states appear. 

 In short,  eq.~(\ref{foundrel1}) follows from the modular properties \cite{Schellekens:1986yi} of certain weak Jacobi forms in terms of which the elliptic genus is expressed \cite{Kawai:1996te,Kawai:1998md,Gritsenko:1999fk}, as will be reviewed in section \ref{subsec_ellgen}.
The mathematical properties of Jacobi forms imply that the sector of the theory charged under the $U(1)$ can
always be characterised by a charge lattice from strings winding on an $S^1$, and it is the physics of such winding states
that governs the charge-to-mass relationship (\ref{foundrel1}) of extremal string states.

\subsection{Content of the present article}

It has already been announced in \cite{Lee:2018urn} that the relation (\ref{foundrel1}) is precisely of the right form in order for the states in the charge sublattice to satisfy a sublattice version \cite{Heidenreich:2016aqi} of the Weak Gravity Conjecture.  To complete the proof of this Sublattice Weak Gravity Conjecture, 
we still need to perform careful calculations of the extremality condition on the black hole side, involving precise numerical factors. 
This is the main purpose of this note. On a side line, we generalize the setup to an arbitrary number of abelian gauge group factors. Along the way we develop a better understanding of the nature of the WGC, which may be useful also beyond the specific applications presented here.

A crucial point is that a proper formulation of the  WGC must address the role of the massless scalar fields that are present in the low-energy effective theory.
There are two, at first sight different ways how such scalar fields enter:
First, a moduli dependence of the gauge kinetic term modifies the charge-to-mass ratio of an extremal black hole \cite{Gibbons:1987ps,Heidenreich:2015nta}; according to one of the original motivations for the WGC \cite{ArkaniHamed:2006dz}, such black holes must be able to decay and hence there must exist particles which are super-extremal with respect to this charge-to-mass ratio.
On the other hand, in \cite{Palti:2017elp} it has been
argued that the particles into which an extremal black hole may decay must not be able to form gravitational bound states. Hence 
the attractive force between two such particles must be smaller than the repulsive gauge forces. If the particles  couple also to massless scalars via a Yukawa interaction, the latter yields an extra contribution to the attractive force in addition to the gravitational force. This leads to an - a priori different  - bound which the particle species must obey \cite{Palti:2017elp}.

As we will argue, if we assume that the WGC is satisfied by a tower of asymptotically massless particles in a neighborhood of the weak gauge coupling point, then the two criteria are automatically equivalent, in this regime of moduli space and in any number of dimensions. 
This is a simple consequence of the exponential collapse of the tower of particle masses near the weak coupling point, as postulated by the Swampland Distance Conjecture. 

Having identified the correct criterion for the WGC particles to fulfill, we carefully examine the role of scalar fields in the 6d F-theory setup introduced in \cite{Lee:2018urn}.
As a result of this analysis we will find that the correct numerical bound to be satisfied is (in units where $M_{\rm Pl} = 1)$
\be \label{SLbound1}
g_{\rm YM}^2 q^2 \stackrel{!}{\geq} \left(\frac{3}{4} + \frac{1}{4}\right) M^2 = M^2 \,.
\ee 
The first term in the brackets is the purely gravitational contribution and the second the scalar-field induced contribution. The latter corresponds to the modification of the extremality condition due to the scalar fields of a dilatonic Reissner-Nordstr{\o}m black hole in the  approach of \cite{Heidenreich:2015nta}, or to the Yukawa force according to \cite{Palti:2017elp}. 
This bound is to be compared with the bound (\ref{foundrel1})  of \cite{Lee:2018urn} for the sublattice (\ref{sublattice1}) of non-BPS states,
\be \label{obsboundintro}
g_{\rm YM}^2 \mathfrak{q}_k^2 = M_k^2 + 4 m g_{\rm YM}^2 \,,
\ee 
which follows from general properties of weak Jacobi forms, as reviewed above.

\begin{figure}[t!]
\centering
\includegraphics[width=10cm] {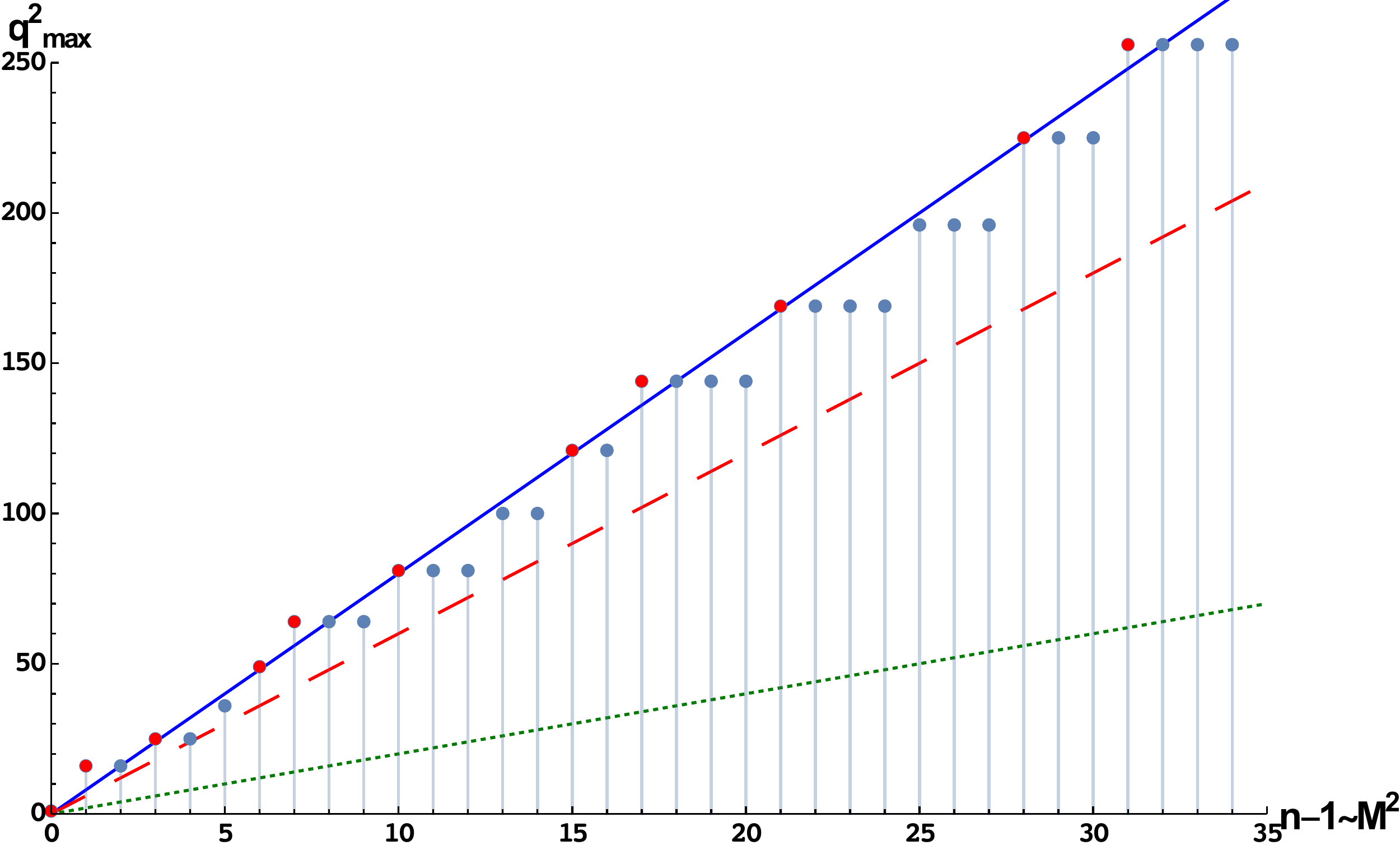}
\caption{ 
Charge-mass spectrum of a 6d  F-theory, or heterotic, string compactification as determined by the elliptic genus,
for a particular example of ref \cite{Lee:2018urn}. Note how narrowly the bound of the Sublattice Weak Gravity Conjecture
(solid blue line) is satisfied by the super-extremal, non-BPS string states (red dots). The new feature we show is how
the blue line arises as the net sum of gravitational (dashed red line) and
scalar (dotted green line) contributions. It is intriguing that this sum, which reflects a zero force property of the physics of extremal (non-BPS) black holes, 
conspires with the spectrum of super-extremal string states, which reflects certain mathematical properties of Jacobi forms.
}
\label{f:ellgenus}
\end{figure}

We see that the sublattice of super-extremal string states bareley lies above the WGC bound (\ref{SLbound1}).  In particular, for large charges $\mathfrak{q}_k$, the additive constant on the RHS of  (\ref{obsboundintro}) can be neglected.  From a mere computational point of view, the outcome that the two relations (\ref{obsboundintro}) and (\ref{SLbound1}) agree so closely may appear as quite miraculous: The relation (\ref{obsboundintro}), or more clearly (\ref{foundrel1}),  is a consequence of the modular properties of the elliptic genus, while (\ref{SLbound1}) arises as the sum of the gravitational and scalar contributions to the zero-force condition of extremal black holes. Viewed in this way, it may seem surprising that the two, a priori different notions of extremality\footnote{Recall that we consider non-BPS, classical black holes here; for BPS states, the equivalence of both notions of extremality is guaranteed by supersymmetry.} give the same numerical bound for the asymptotic charge-to-mass ratio. 
See Fig.~\ref{f:ellgenus} for a suggestive visualisation. 

The physics behind this seeming coincidence, is, of course,  analogous to the behaviour observed already for the heterotic string on flat backgrounds in \cite{ArkaniHamed:2006dz,Heidenreich:2016aqi}: It reflects the fact that the highly excited string states can asymptotically become extremal black holes themselves \cite{Sen:1995in}.

Suppose the theory gave, instead of (\ref{obsboundintro}), a relation of the form  $g_{\rm YM}^2 \mathfrak{q}_k^2 = c M_k^2 + \ldots$ with $c > 1$ for a set of physical states, and with corrections which become subleading for high charges.
If the WGC is satisfied by a sublattice of states, 
 the constant $c$ \emph{must} 
be given by $c=1$:
Otherwise, for high charges we would form a black hole whose charge-to-mass ratio exceeds the extremality bound (\ref{SLbound1}). 
In this sense there are only two options: Either violate the sublattice version of the WGC altogether, or satisfy it {\it \`a point}. This principle seems to be deeply built into string theory, and in particular cannot depend on any BPS property.

As we will detail further, while for the tower states the Coulomb interaction exceeds the combined attraction of gravity and scalar interactions, in the spirit of the WGC, the gravitational force as such is not weaker, but stronger compared  to the scalar force acting between two such test particles. The slogan `gravity is the weakest force' hence holds only with a grain of salt, at least for the tower of WGC states which become asymptotically massless near weak coupling in the present setup. 

In a secondary direction of this article, we generalize the above findings to an arbitrary number of abelian gauge group factors: 
At a technical level, the elliptic genus of the asymptotically tensionless string now involves a lattice quasi-Jacobi form of higher rank \cite{Ziegler}:  Its elliptic transformation behaviour is determined by a {\it matrix} which underlies the definition of a higher-dimensional charge lattice. Higher rank Jacobi forms comprise, as a special case, Weyl invariant Jacobi forms, whose associated charge lattices are those of simple Lie algebras.
Such generalizations of weak and of Weyl invariant Jacobi forms are an active field of arithmetic geometry \cite{Zagiertoappear}. While to date an explicit basis of the set of generators, comparable to the rank one case, is not available in the literature, the theta-decomposition already invoked in \cite{Lee:2018urn} to prove (\ref{sublattice1}) carries over to several $U(1)$s. To confirm this general pattern in an example, we will propose a closed form for the elliptic genus in an F-theory model with two abelian gauge factors, which passes non-trivial consistency checks against the BPS invariants as computed by mirror symmetry.  It would be interesting to explore the application of higher rank quasi-modular forms along these lines further.

More specifically, the present article is structured as follows: In section \ref{subsec_super} we contrast the two a priori different ways how massless scalar fields enter the WGC, by modifying the extremality bound \cite{Heidenreich:2015nta} or by contributing extra scalar forces \cite{Palti:2017elp}.
In section \ref{sec_SWCG} we point out under which assumptions the resulting WGC bounds are guaranteed to agree. 
In section \ref{subsec_WGCinFa} we prove the WGC in the most general F-theory compactification to 6d with 8 supercharges near a weak coupling point. 
After  setting the stage of the relevant chiral supergravity theory in section \ref{subsec_WGCinFb}, we carefully compute, in section \ref{subsec_weaklimit}, the scalar field kinetic metric in the weak coupling limit introduced in \cite{Lee:2018urn}.  This allows us to prove that the Sublattice WGC conjecture is indeed satisfied in the weak coupling regime in section \ref{subsec_proving}.
The discussion is phrased directly for several abelian gauge fields. In section \ref{subsec_ellgen} we first prove the generalization of the relation (\ref{foundrel1}) from  \cite{Lee:2018urn}  for multiple abelian gauge fields, by invoking a theta decomposition of higher-rank quasi-Jacobi forms.
Section \ref{subsec_example} demonstrates this in an explicit example, which also shows how to overcome the limitations in the literature in providing an explicit basis of such forms.

\section{\large On the Scalar Weak Gravity Conjecture}

The main aim of this work is to verify the Weak Gravity Conjecture (WGC) in the presence of massless scalar fields in a concrete string theoretic setting. 
As described in the Introduction, there exist seemingly different formulations and justifications for this conjecture, and it may a priori not be obvious to what extent they are strictly equivalent. In order for the comparison to be meaningful at a quantitative level, it is important to keep track of the precise numerical factors that appear in each formulation.
For these reasons we now revisit the WGC in the presence of massless scalar fields. 

\subsection{Super-extremality versus balancing of forces} \label{subsec_super}

According to one of the original arguments in \cite{ArkaniHamed:2006dz} for motivating the WGC bound, charged extremal black holes must be able to decay.
In absence of massless scalar fields the extremal black holes charged under the abelian gauge symmetry  correspond to extremal Reissner-Nordstr{\o}m black holes.
In
 the context of a supergravity theory, on the other hand, the couplings of the effective theory are controlled by the vacuum expectation value of moduli fields.
 If the latter are massless, a charged extremal black hole solution imprints a non-trivial profile of the scalars entering the gauge coupling. This scalar profile modifies the charge-to-mass ratio of an extremal charged black hole, and it is this modified value that constitutes the lower bound for the charge-to-mass ratio of a particle satisfying the WGC \cite{Heidenreich:2015nta}. 
 
 Let us consider  a theory with several abelian gauge group factors $U(1)_a$, $a = 1, \ldots, n_{U(1)}$, whose kinetic terms depend on a set of scalar fields $\phi^r$, $r = 1, \ldots n_s$. This is encoded in the following action:
 \be \label{SEMd}
S_{\rm EMd} = \int_{\mathbb R^{1,d-1}}      \frac{M^{d-2}_{\rm Pl}}{2}  \sqrt{-g}    R - \frac{M^{d-2}_{\rm Pl}}{2} g_{r s} {\rm d} \phi^r \wedge \ast {\rm d} \phi^s -    \frac{1}{2}  f_{ab}(\phi^r)  \,  
 F^a \wedge \ast F^b \,.
\ee
The $d$-dimensional Planck mass $M_{\rm Pl}$ has been included in such a way that the scalars, which we write in an obvious vector notation as $\vec{\phi}$, have mass dimension zero.
We are working in a frame where the scalar kinetic matrix $g_{r s}$ is taken to be constant.

As a special case the gauge kinetic terms can depend exponentially on $\vec{\phi}$. Such a dilatonic theory is hence characterized by a matrix of
 gauge kinetic terms
\be \label{gphicoupling}
 f_{ab}(\vec\phi)  =  f_{ab}(\vec{\phi}_0)  \,  e^{  \vec{\alpha} \cdot (\vec{\phi} - \vec{\phi}_0)} \ 
\ee
in terms of the invariant combination
\be
\vec{\alpha} \cdot \vec{\phi} = g_{r s}  \alpha^r \phi^s \,.
\ee

In this situation and with the above normalization of the fields, the charge-to-mass ratio of an extremal dilatonic Reissner-Nordstr{\o}m black hole with total charges $Q_{a}$ and ADM mass $M_{\rm ADM}$ is given by \cite{Gibbons:1987ps,Duff:1996hp}
\be \label{WGC1}
\langle\underline Q, \underline Q\rangle_{f(\vec\phi_0)} :=Q_a f^{ab}(\vec{\phi}_0) Q_b = \mu \frac{M_{\rm ADM}^2}{M^{d-2}_{\rm Pl}}  \,, \qquad
\ee  
with
\be
\begin{split}
 \mu = \mu_G + \mu_{\vec{\phi}} \,, \qquad 
\mu_G = \frac{d-3}{d-2} \,, \qquad \mu_{\vec{\phi}} = \frac{1}{4} \vec{\alpha} \cdot \vec{\alpha}   \label{muGmuphi} \,.
\end{split}
\ee
{Here we have identified the reference value $\vec{\phi}_0$ with the asymptotic value of the scalar fields $\vec\phi$ in the extremal dilatonic Reissner-Nordstr{\o}m black hole solution,
 \be
 \vec\phi \to \vec{\phi}_0  \qquad {\rm as}  \quad  r \to \infty \,,
 \ee
 where $r$ is the spatial radial coordinate $r$.
 It is this asymptotic value which controls the  extremality condition (\ref{WGC1}), where
 $f^{ab}(\vec{\phi}) = f^{ab}(\vec{\phi_0}) e^{-\vec{\alpha} \cdot (\vec{\phi}-\vec{\phi_0})}$ is the inverse of the gauge kinetic matrix $f_{ab}(\vec{\phi})$.
Note that if the scalars $\vec{\phi}$ were massive and hence not activated in the black hole solution, the dilatonic contribution $ \mu_{\vec{\phi}}$ would be absent. In this sense, $\mu_G$ represents the purely gravitational contribution to the extremality condition.

The WGC then requires the existence of a suitable set of particles of charges $\underline{\mathfrak{q}}$ 
and associated masses $\mathfrak{m}$ whose charge-to-mass ratio exceeds this extremality bound~\cite{Heidenreich:2015nta} in such a way that every extremal charged dilatonic black hole can decay into super-extremal particles  \cite{ArkaniHamed:2006dz}.
The minimal requirement which guarantees this in the presence of several $U(1)$ factors is the 
  Convex Hull Condition of \cite{Cheung:2014vva}. 
 A stronger condition, which is the one we will be studying in this article, is the Sublattice Weak Gravity Conjecture \cite{Heidenreich:2016aqi}. It postulates that (at least) a sublattice of the charge lattice is populated by physical states whose charge-to-mass ratio exceeds the extremality bound (\ref{WGC1}),
\be \label{WGCgeneral1}
 \langle \underline{\mathfrak{q}}, \underline{\mathfrak{q}} \rangle_{f(\vec\phi)}    \stackrel{!}{\geq} \mu \, \frac{\mathfrak{m}^2}{M^{d-2}_{\rm Pl}} \,.
\ee
This condition implies the Convex Hull Condition of \cite{Cheung:2014vva} and moreover is stable under dimensional reduction \cite{Heidenreich:2015nta}. 
Note that for a spatially varying scalar field configuration, we immediately face the question which specific value of $\vec{\phi}$ appears in (\ref{WGCgeneral1}), i.e. for which values of $r$ the equation is to be evaluated \cite{Klaewer:2016kiy}. 
A related issue is whether also the mass $\mathfrak{m}$ of the test particle on the RHS is field dependent. We will address both points in section \ref{sec_SWCG}.

Let us now turn to an at first sight rather different interpretation of the  WGC, which posits that `gravity should be the weakest force' \cite{ArkaniHamed:2006dz}. This means that there must exist a set of particles (possibly populating a charge sublattice)
such that for two such test particles the attractive gravitational interaction is smaller than the repulsive Coulomb interaction. 
When extra massless scalar fields are present, ref.~\cite{Palti:2017elp}  has proposed to modify this criterion such that for two such test particles,  the following stronger force balancing condition 
\be \label{Vrequirement}
|F_{\rm Coulomb}| \stackrel{!}{\geq }|F_{\rm Newton}|  + |F_{\rm Yukawa}| \,
\ee
must be imposed.
Here $F_{\rm Coulomb}$ and $F_{\rm Newton}$  describe the classical electromagnetic and gravitational force  between two test particles of mass $\mathfrak{m}$ and $U(1)_a$ charges $\mathfrak{q}_a$.
The crucial insight of \cite{Palti:2017elp} is to include the attractive force $F_{\rm Yukawa}$ between two test particles that couple to a set of massless scalar fields $\Phi^i$ via a Yukawa interaction. 
Among other things it has been shown in \cite{Palti:2017elp} that BPS states in 4d ${\cal N}=2$ supergravity necessarily
saturate the inequality because BPS objects do not exert any net force onto each other.

To compare the two formulations (\ref{WGCgeneral1}) and (\ref{Vrequirement}) of the WGC, in an arbitrary number of dimensions and also for non-BPS particles, consider the action 
\bea
S &=& S_{\rm EMd} + S_{\rm Yuk} \\
S_{\rm Yuk} &=&  \int_{\mathbb R^{1,d-1}}   -  \frac{M^{d-2}_{\rm Pl}}{2}   g_{ij} {\rm d} \Phi^i \wedge \ast {\rm d} \Phi^j +  \bar \psi  i \gamma \cdot (\partial + i \mathfrak{q}_a A^a) \psi + \mathfrak{m}(\vec{\Phi}) \bar \psi \psi  \,.
\eea
Here $\psi$ denotes a charged test particle, which is taken to be fermionic for definiteness. It couples not only to the gauge fields $A^a$ (whose dynamics is described by $S_{\rm EMd}$ as in (\ref{SEMd})), but in addition to some massless real scalars 
\be
\Phi^i = \varphi^i + \delta \varphi^i \,,
\ee
where by $\delta \varphi^i$ we are denoting the fluctuations around the background configuration $\varphi^i$.
With this split we can write
\be
\mathfrak{m}(\vec{\Phi}) = \mathfrak{m}(\vec{\varphi}) + \frac{\partial \mathfrak{m}(\vec{\varphi})}{\partial \varphi^i} \, {\delta\varphi}^i    + \ldots
\ee
 so that the mass of the test particle,  $\mathfrak{m}(\vec{\varphi})$, is controlled by the background value $\vec{\varphi}$ of the dimensionless massless scalars. 
 Similarly, 
\be
h_i(\vec\varphi) := \frac{\partial \mathfrak{m}(\vec\varphi)}{\partial \varphi^i}  \equiv  \partial_i \mathfrak{m} (\vec{\varphi})
\ee
governs a triple Yukawa interaction 
\be
S_{\rm Yuk} \supset \int_{\mathbb R^{1,d-1}} h_i(\vec{\varphi}) \, \delta\varphi^i \,  \bar \psi \psi  \,.
\ee
Note that a priori the scalars $\varphi^i$ which determine the mass of the test particle and the scalars $\phi^r$ which control the gauge kinetic matrix as in (\ref{SEMd}) are independent.\footnote{In this sense, it is a slight abuse of notation to assemble both types of fields into vectors denoted by the same vector-symbol. If there is a (partial) overlap between the set of scalar fields $\phi^r$ and $\Phi^i$, it is understood that the kinetic terms appear only once of course.} In fact, from a purely low-energy effective field theory point of view,
a dependence of the test particle mass on scalar fields, as encoded in $S_{\rm Yuk}$, would be perfectly consistent with the absence of any dilaton field in the gauge kinetic term.

The Coulomb, gravitational, and Yukawa forces between two test particles of $U(1)_a$ charges $\mathfrak{q}_a$ and mass $\mathfrak{m}(\vec{\varphi})$ then take the textbook form
\be
|F| = \frac{A}{{\rm Vol}(S^{d-2}) \, r^{d-2}} 
\ee
with
\bea
A_{\rm Coulomb} =  \langle \underline{\mathfrak{q}},  \underline{\mathfrak{q}} \rangle_{f(\vec{\phi})}  \,, \qquad  
A_{\rm Newton} =   \frac{\mathfrak{m}(\vec{\varphi})^2}{M^{d-2}_{\rm Pl}}\,  \frac{d-3}{d-2}, \qquad 
A_{\rm Yukawa} =    \frac{h_i(\vec{\varphi}) \, g^{ij}\,  h_j(\vec{\varphi})}{M^{d-2}_{\rm Pl}} \,.
\eea
Here ${\rm Vol}(S^{d-2})$ is the volume of the unit sphere in $d-1$ spatial dimensions. 
The gravitational force has a dimension-dependent prefactor, which has been carefully derived for instance in \cite{Robinson:2006yd}. It originates in taking the non-relativistic limit of the Einstein equations in $d$ dimensions and has no
analogue for the Coulomb and Yukawa force. Note furthermore that the coupling appearing in $A_{\rm Coulomb}$ involves the inverse of the background gauge kinetic matrix, evaluated at a certain reference value $\vec{\phi}$ to be discussed momentarily.

The requirement (\ref{Vrequirement}) for two test particles of species $\psi$ then translates into the constraint
\be  \label{WGC2}
     \langle  \underline{\mathfrak{q}},  \underline{\mathfrak{q}} \rangle_{f(\vec{\phi})}    \stackrel{!}{\geq}  \frac{ \mathfrak{m}(\vec{\varphi})^2}{M^{d-2}_{\rm Pl}} \left(  \frac{d-3}{d-2} \,+  \frac{1}{\mathfrak{m}(\vec{\varphi})^2} \,  g^{ij} \partial_i \mathfrak{m}(\vec{\varphi}) \partial_j \mathfrak{m}(\vec{\varphi})    \right) \,.
\ee
This formula generalises the expression postulated in \cite{Palti:2017elp} for $d=4$, which was shown to be satisfied by charged BPS states in ${\cal N}=2$ supergravity.

\subsection{The Scalar Weak Gravity Conjecture near a weak coupling point} \label{sec_SWCG}

Comparing the two constraints (\ref{WGCgeneral1}) and (\ref{WGC2}),
we first observe that the numerical prefactors multiplying $\mathfrak{m}^2$  in the purely gravitational parts - encoded in $\mu_G$ and $A_{\rm Newton}$, respectively - perfectly agree.
On the other hand, the scalar contributions - $\mu_\phi$ and $A_{\rm Yukawa}$ - 
 can only be equivalent if the moduli dependence of the gauge coupling  and the moduli dependence of the mass term of the test particle 
are properly correlated. 

It is conceivable that the strict equivalence of (\ref{WGCgeneral1}) and (\ref{WGC2}) can  be argued for by black hole decay arguments: If an extremal black hole is supposed to decay into a set of particles which do not satisfy (\ref{WGC2}), the decay products could form bound states as the attractive force exceeds the repulsion. This was discussed, along with other aspects of the criterion (\ref{WGC2}), in ref. \cite{Palti:2017elp}, which motivated (\ref{WGC2}) by requiring that the decay products of an extremal black hole should not be able to form  gravitational bound states. As it stands,  this argument just posits that there must exist particles satisfying  (\ref{WGCgeneral1}) and in addition (\ref{WGC2}), and in principle there is room for a numerical difference between these two criteria.

Our approach to understanding the relation between (\ref{WGCgeneral1}) and (\ref{WGC2}) follows a different line of arguments: Suppose a particle satisfies either of the WGC bounds at a given point in moduli space. By continuity it is reasonable to expect that it continues to satisfy the bound in an entire  neighborhood.
This principle was dubbed `Local Weak Gravity Conjecture' in \cite{Klaewer:2016kiy}, which studies in detail the relation between a field dependent version of the WGC and the Swampland Distance Conjecture.
Clearly, in such a neighborhood, the moduli dependence of the bilinear form $ \langle \mathfrak{q}, \mathfrak{q} \rangle_f = \mathfrak{q_a} f^{ab}  \mathfrak{q_b}$ and of $\mathfrak{m}$ must be related for the WGC particle.
In fact, if we assume that the WGC particles populate (at least) a charge sublattice,
 the moduli dependence of $\mathfrak{m}(\vec{\varphi})$ for the WGC particles and of the inverse gauge kinetic matrix $f^{ab}(\vec{\phi})$ must agree in the neighborhood where the WGC is satisfied. 
The reason is that for asymptotically large charges the super-extremal particles must {\it saturate} the extremality bound. Hence, the functional dependence of both sides of the extremality bound on the scalar fields must be the same. 

This in particular identifies the moduli controlling the mass  of the WGC particles with the moduli entering the gauge coupling, and we conclude that
\be \label{modulidependence1}
f^{ab}(\vec\varphi) = f^{ab} \,  F(\vec\varphi) \,,  \qquad \mathfrak{m}^2(\vec\varphi) = \mathfrak{m}^2  \, F(\vec\varphi)
\ee
in a neighborhood around the point  in moduli space where the particles in question satisfy the WGC bound.

Suppose now furthermore that we are considering the limit in moduli space where 
\be \label{weakcouplingseveral}
f^{ab}(\vec\varphi) \to 0 \,,\quad M_{\rm Pl} \quad {\rm finite}   \qquad {\rm as} \quad   \vec\varphi \to \vec{\varphi}_\infty \,.
\ee
This is what is meant by the `weak coupling regime' in moduli space for several abelian factors: In particular, all eigenvalues of $f^{ab}(\vec\varphi)$ must approach zero\footnote{In principle only a subset of the eigenvalues may approach zero, in which case we have to restrict to the corresponding subset of $U(1)$s. See the discussion after eq.~\eqref{detmcond}.} and there must be an overall asymptotic behaviour of the form (\ref{weakcouplingseveral}).

In order to avoid tension with the conjectural
requirement that no global continuous  symmetries exist in any consistent theory of quantum gravity \cite{Banks:1988yz,Banks:2010zn}, the weak coupling point $\vec{\varphi}_\infty$ must lie at infinite distance in moduli space. 
According to the Swampland Distance Conjecture \cite{Ooguri:2006in}, a tower of states becomes massless whenever a point at infinite distance is approached, and the mass is conjectured to decrease exponentially in the distance in moduli space.
 This means that the mass scale of the particle tower should asymptotically behave like
\be
\mathfrak{m}^2(\vec\varphi) = \mathfrak{m}^2 \, e^{- \vec{c}  \cdot \vec\varphi }  \qquad  {\rm as} \quad   \vec\varphi \to \vec{\varphi}_\infty
\ee
for some order one numbers $c_i$. (Recall that we are working in conventions where the scalar fields $\varphi^i$ are dimensionless.) The fact that the $c_i$ are of order one is the content of the Refined Swampland Distance Conjecture \cite{Klaewer:2016kiy}, but this plays no essential role for us here. 
Suppose furthermore that the tower of asymptotically massless states satisfies the WGC. Then in a neighborhood of the point at infinity, we must identify the moduli dependence appearing in (\ref{modulidependence1}) as  
\be
F(\vec\varphi) = e^{- \vec{c} \cdot \vec\varphi } \qquad  {\rm as} \quad   \vec\varphi \to \vec{\varphi}_\infty\,.
\ee
This inevitably puts us in the context of dealing with a dilatonic type black hole with coupling (\ref{gphicoupling}) with 
\be
\vec\alpha =   \vec{c} \,,
\ee
at least in the neighborhood of the weak coupling point.
With this input, the two criteria (\ref{WGCgeneral1}) and (\ref{WGC2}), including the precise numerical factors, for the asymptotically massless WGC particles are equivalent by inspection, in the vicinity of the weak coupling point.\footnote{The fact that two extremal (non-SUSY) dilatonic Reissner-Norstr{\o}m black holes exert no net force on each other (`equipoise' or `anti-gravity') is a classic result \cite{Gibbons:1987ps, Gibbons:1993dq}.}
In particular, the super-extremality condition yields the same constraint at every value of the scalar fields in the weak coupling region, simply because the field dependence cancels on both sides of the equation. 
Of course, this was essentially put in by assuming that the WGC holds not just at isolated points, but in a region of moduli space \cite{Klaewer:2016kiy}.

Let us summarize this discussion in the form of the following \\

\noindent {\bf Observation}:
{\it Consider the point $\vec{\varphi}_\infty$ at infinite distance in moduli space where the gauge coupling matrix $f^{ab}(\vec\varphi)$ vanishes asymptotically. By the Swampland Distance Conjecture a tower of states becomes massless with $\mathfrak{m}^2(\varphi) = \mathfrak{m}^2 \, e^{-\vec{c} \cdot \vec{\varphi} }$ as $\vec\varphi \to \vec{\varphi}_\infty$.
If a subset of these asymptotically satisfies the Sublattice Scalar Weak Gravity Conjecture in the form (\ref{WGC2}) in a neighborhood of $\vec{\varphi}_\infty$, then at least  in this neighborhood also $f^{ab}(\vec\varphi) = f^{ab} e^{-\vec{c}\cdot \vec{\varphi} }$, and the criterion (\ref{WGC2}) is equivalent to the super-extremality condition (\ref{WGCgeneral1}) at any point in the neighborhood. \\
}

We believe that this is indeed the general behaviour in the vicinity of an asymptotic weak coupling point, as summarized in the following  \\

\noindent{\bf Conjecture}:
{\it Near a weak coupling point (\ref{weakcouplingseveral}), the charge vectors of a tower of asymptotically massless states span at least a sublattice of the full charge lattice, and the corresponding physical states satisfy the Sublattice Weak Gravity Conjecture bound (\ref{WGC2}) at least in a neighborhood of the weak coupling point. By the above Observation,  (\ref{WGC2}) is then guaranteed to be equivalent to super-extremality, (\ref{WGCgeneral1}), with respect to a dilatonic extremal charged black hole.} \\

It is interesting to speculate what happens away from the weak coupling point. 
A priori, it is not clear if the Weak Gravity Conjecture must be satisfied at an arbitrary point in moduli space, in particular at those where the gauge couplings are far from zero.\footnote{We thank Irene Valenzuela and Eran Palti for interesting discussions  on this.} 
Verfiying the equivalence between the super-extremality and the force balancing condition in such regimes requires knowledge of the extremal black hole solution for more general gauge kinetic matrices.
Turning tables round, if the two criteria were guaranteed to be equivalent away from weak coupling on general grounds, then this would give a much simpler way to compute the extremality bound of a Reissner-Nordstr{\o}m black hole with scalar charge for an arbitrary moduli dependence of the coupling.  Clearly this should be understood better.

As a final remark, let us compare the purely gravitational and the scalar field induced terms on the RHS of (\ref{WGC2}), in the regime where $\mathfrak{m}(\varphi) = \mathfrak{m} \,  e^{-\frac{1}{2}\vec{c}\cdot \vec\varphi   }$.
Their relative size depends on the ratio between $\frac{d-3}{d-2}$ and $\vec{c}^2/4$.  We will see in this paper that there are consistent string constructions in which the gravitational interaction is in fact stronger than the Yukawa interaction in the sense that 
\be
\frac{d-3}{d-2} > \frac{\vec{c}^2}{4}
\ee
 for the tower of states satisfying the WGC near the weak coupling point. 
Of course, there may exist other particles for which the Yukawa interaction is smaller than the gravitational one. Whether such particles always exist, in agreement with another conjecture in \cite{Palti:2017elp}, would be interesting to investigate further. 

\section{\large The Scalar Weak Gravity Conjecture in F-theory} \label{subsec_WGCinFa}

In this section we prove the WGC in the vicinity of a weak coupling point in 6d F-theory compactifications with 8 supercharges. By duality the result carries over to 6d compactifications of the heterotic string with the same amount of supersymmetry.
As we have seen, the precise formulation of the WGC heavily depends on the dynamics of the light scalar fields in the theory.
In order to study their effect at a quantitative level, we first translate the parametrization of the weak coupling limit given in \cite{Lee:2018urn} into the formalism of 6d $N=(1,0)$ supergravity, in which the scalar kinetic terms have been reliably computed in the literature. This allows us to evaluate the WGC bound in this context and to compare it to the charge-to-mass ratio of the tower of asymptotically massless states found in \cite{Lee:2018urn}.

\subsection{General form of the effective action} \label{subsec_WGCinFb}

The low-energy effective theory of F-theory compactified on an elliptic Calabi-Yau three-fold $Y_3$ is described by a 6d $N=(1,0)$ supergravity. This chiral theory
contains $n_T$ tensor multiplets, whose bosonic components comprise an anti-self-dual 2-form field along with a real scalar. 
Including the self-dual gravi-tensor we denote the 2-form fields as $B^\alpha$, $\alpha = 0, \ldots, n_T$, and the real scalar fields as $j^\alpha$. The latter are subject to the constraint
\be
j \cdot j = \Omega_{\alpha \beta} j^\alpha j^\beta  = 1 \,,
\ee
where $\Omega_{\alpha \beta}$ is an $SO(1,T)$ invariant inner product. Without loss of generality we can go to a diagonal basis where 
\be \label{diagonalOmega}
\Omega_{\alpha \beta} = {\rm diag}(1,-1, \ldots, -1) \,.
\ee
The pseudo-action of the $6$d $\cN=(1,0)$ supergravity theory coupled to $n_V$ abelian gauge fields has been worked out in \cite{Ferrara:1997gh}. In the notation of \cite{Bonetti:2011mw} it can be written as
\beq\label{pa1}
S= \int_{\IR^{1,5}} \frac12 R *1 -\frac12 g_{\alpha\beta}{\rm  d}j^\alpha \wedge *{\rm d}j^\beta- \frac{1}{2} (j\cdot b_{ab}) \, F^a\wedge *F^b  + S_{\rm tensor} + S_{\rm matter} \,.
\eeq
Here the kinetic metric is defined as
\beq\label{met}
g_{\alpha \beta} = 2j_\alpha j_\beta - \Omega_{\alpha \beta}\,,
\eeq 
and we have set the 6-dimensional Planck mass to unity.\footnote{If we generalise to non-abelian gauge groups, we must replace $F^a \wedge \ast F^b$ by $\frac{1}{\lambda_I} {\rm tr} F_I \wedge \ast F_I$, where $\lambda_I$ is the Dynkin label associated with the Lie algebra and takes the value $\lambda =1$ for gauge algebra $\mathfrak{su}(N)$.} Furthermore the gauge kinetic terms are controlled by the parameters $b_{ab}^\alpha$.
The tensor part of the pseudo-action takes the form
\be \label{patensor}
S_{\rm tensor} = -\frac{1}{4} g_{\alpha \beta} G^\alpha \wedge \ast G^\beta - \frac{1}{2} \Omega_{\alpha \beta} B^\alpha \wedge X_4^\beta \,.
\ee
The gauge invariant 3-form field strength
\be
G^\alpha = d B^\alpha + \frac{1}{2} a^\alpha \omega_{\rm L} + 2 b^\alpha_{ab} \omega^{a b}_{\rm YM} 
\ee
is subject to the self-duality constraint
\be
g_{\alpha \beta} \ast G^\beta = \Omega_{\alpha \beta} G^\beta \,,
\ee
which is to be imposed at the level of the equations of motion.
The Chern-Simons terms are defined in the usual way such that the Bianchi identity takes the form
\be
d G^\alpha  = \frac{1}{2} a^\alpha {\rm tr} R^2 + 2 b_{a b}^\alpha  F^a \wedge F^b \,.
\ee
Finally, $S_{\rm matter}$ contains the hypermultiplets of the theory, which includes the sector of charged matter fields. 

An important detail for us is that we are normalizing the gauge kinetic term as in (\ref{pa1}), which differs by a relative factor of $\frac{1}{4}$ from the conventions of \cite{Ferrara:1997gh,Bonetti:2011mw}. This normalization, together with the definition of the tensor field pseudo-action (\ref{patensor}), implies that in particular the quartic  anomaly equation for ${U(1)_a}^4$ takes the form
\be \label{anomalyequation}
\frac{1}{3} \sum_n N_n (\mathfrak{q}^{(n)}_a)^4 = b_{a a} \cdot b_{a a} \,.
\ee
Here, $\mathfrak{q}_a^{(n)}$ denotes the $U(1)_a$ charge of the canonically normalised massless hypermultiplet fermions and $N_n$ is the number of such hypermultiplets.
In particular, the charge vector $\mathfrak{q}_a$ appears in the covariant derivative $D_\mu = \partial_\mu + i \mathfrak{q}_a A^a_\mu$ acting on the charged matter hypermultiplets.

The above quantities are related to the geometric data of an F-theory compactification on an elliptic three-fold $Y_3$ with base $B_2$ as follows\footnote{The effective action of 6d F-theory compactifications has been under intense investigation in the recent literature, including \cite{Kumar:2009ac,Taylor:2011wt,Park:2011wv,Bonetti:2011mw}. Background on F-theory and the realisation of abelian gauge group factors can be found in the recent reviews \cite{Weigand:2018rez,Cvetic:2018bni}, to which we also refer for  a more complete account of the original literature.}:
The object $\Omega_{\alpha \beta}$ can be interpreted as the topological intersection form of a basis of $H^{1,1}(B_2, \mathbb R)$.
In particular we can introduce a basis $\{w_\alpha\} = \{w_0, w_i\}$ of $H^{1,1}(B_2, \IR)$ such that $\Omega_{\alpha \beta} = \omega_\alpha \cdot \omega_\beta$ is in the diagonal form (\ref{diagonalOmega}).
Given such a basis, the scalar fields $j^\alpha$ are interpreted as the components of the object $j = j^\alpha \omega_\alpha$ which is related to the
 K\"ahler form $J$ of $B_2$ via
\be
J = \sqrt{2 {\cal V}} \,  j \,,\qquad \qquad  {\cal V} = {\rm vol}(B_2) \,.
\ee
The normalization $j \cdot j =1$ then implements that ${\rm vol}(B_2) = \frac{1}{2} J \cdot J$, as required. 
The real scalar degrees of freedom can always be organized into a distinguished real scalar $x$ and a set of $n_T-1$ real scalars $\phi^a$, $a = 1, \ldots, n_T-1$, by parametrizing  
\bea \label{parametrisation}
j^0 &=& {\rm cosh}(x)\,,\quad j^i = u^i(\phi^a)\,{\rm sinh}(x)\,,\quad i=1, \dots, n_T\,.
\eea
Here $u^i$ are functions of $\phi^a$ with 
\be
\sum\limits_{i=1}^{n_T} (u^i)^2 = 1.
\ee
Indeed, this is the most general parametrization for $j$ subject to $j \cdot j =1$. 
The kinetic matrix (\ref{met}) in this basis becomes
\bea
g_{\alpha \beta}
&=&  \left (\begin{array}{cc} {\rm cosh} (2x) & - {\rm sinh} (2x) \,u^j\\ - {\rm sinh} (2x) \, u^i& 2{\rm sinh}^2(x) \,u^iu^j  + \delta^{ij} \end{array}\right )\,,
\eea
from which the kinetic terms for the independent real scalar fields $x$ and $\phi^a$ are computed as 
\beq \label{kinetictermx}
\frac12 g_{\alpha \beta}  {\rm d}j^\alpha \wedge * {\rm d} j^\beta = \frac12 {\rm d}x\wedge *{\rm d} x + {\rm sinh}^2(x) \sum_{i=1}^{n_T} {\rm d}u^i(\phi^a) \wedge *{\rm d}u^i(\phi^a) \,.
\eeq
Finally, the anomaly coefficients $b_{ab}^\alpha$, which control both the inverse gauge coupling and the form of the anomaly counterterms, are the coefficients of a two-cycle class $C_{ab}$,
\be
C_{ab} = b_{ab}^\alpha \, \omega_\alpha \,.
\ee
Here  \cite{Park:2011ji}
\be
C_{ab} = -\pi_\ast(\sigma(S_a) \cdot \sigma(S_b))
\ee
 is the height pairing matrix associated with the rational sections $S_a$ on the elliptic fibration $Y_3$, which underlie the definition of the abelian gauge group factors in F-theory.
 With these conventions, the $U(1)_a$ charges $\mathfrak{q}_a$ are integral, and 
it can be checked that this definition is consistent with the anomaly equation (\ref{anomalyequation}).
In particular, the gauge kinetic matrix is determined as
\be \label{fab6d}
f_{ab} = j \cdot b_{ab} = \frac{1}{\sqrt{2 {\cal V}}}  J \cdot C_{ab} \,.
\ee

\subsection{The weak coupling limit for several U(1)s} \label{subsec_weaklimit}

We now wish to parametrise the limit in moduli space where the $n_V$ abelian gauge factors become asymptotically weakly coupled, while at the same
time the 6d Planck mass stays finite.
The special case of a single abelian gauge group factor has already been treated in detail in \cite{Lee:2018urn}: In this case,
the inverse gauge coupling is controlled by the volume of a single curve $C$ on $B_2$.
In order for this volume to become infinite, while the volume of $B_2$ stays finite, the K\"ahler form $J$ of the F-theory base $B_2$ must asymptotically take the form 
\be \label{Jasymptotic1}
J = t J_0 +  s_\nu I_\nu \,.
\ee
Here $J_0$ and $I_\nu$ are the integral classes that generate the K\"ahler cone of $B_2$ and $J_0$ has the property that $J_0 \cdot J_0 = 0$. Furthermore
\be
m = \frac{1}{2} C \cdot J_0
\ee
is non-vanishing. As a result, the volume of the curve $C$ which controls the inverse gauge coupling diverges as $t \to \infty$.
A careful analysis of the ansatz  (\ref{Jasymptotic1}) in terms of the K\"ahler cone parameters shows that $s_\nu$ must vanish in this limit as\footnote{In \cite{Lee:2018urn} we had set ${\cal V} \equiv 1$, but there is no harm in leaving ${\cal V}$ general, but fixed at a finite value.} 
\be \label{1overtconstraint}
\sum_\nu s_\nu (J_0 \cdot I_\nu) \to  {\cal V}/t
\ee
in order for the volume of $B_2$ to stay finite.
For a given base $B_2$ and curve $C$ it may not be possible to take such a limit. In this case, the infinitesimal coupling point cannot be taken geometrically while keeping the volume of $B_2$ finite. If such a limit exists, on the other hand, it must be of the above form.

In the presence of several abelian gauge group factors, this discussion generalizes in that we now require that all eigenvalues of the gauge kinetic matrix $f_{ab}$ in (\ref{fab6d}) asymptote to infinity,  while the volume of $B_2$ stays finite.  That is, we need to consider a limit where the inverse of the gauge kinetic matrix behaves as 
\be \label{JCabinft}
f^{ab} =\sqrt{2 \cal V} (J \cdot C^{ab}) \to 0 \,.
\ee
Indeed, this ensures that $f^{ab} \mathfrak{q}_a \, \mathfrak{q}_b \to 0$ for any charge vector $\mathfrak{q}_a$, which is the combination that controls the gauge interactions as discussed in the previous section. 
The requirement (\ref{JCabinft}) simply means that $J_0$ in (\ref{Jasymptotic1}) must be chosen in such a way that the matrix
\be \label{mabdef}
m_{ab} = \frac{1}{2} C_{ab} \cdot J_0
\ee
is of maximal rank, i.e.
\be \label{detmcond}
{\rm det}(m_{ab}) \neq 0 \,.
\ee
Again, in a given geometry with a specific choice of gauge kinetic matrix (\ref{fab6d}), it may not be possible to take such a limit (\ref{Jasymptotic1}) subject to (\ref{detmcond}). In this case, the weak coupling limit can be taken only for a (possibly empty) subset of the abelian gauge group factors. In the sequel, we focus on those abelian gauge factors for which (\ref{detmcond}) holds, and study the Weak Gravity Conjecture for this kind of gauge groups in the vicinity of the weak coupling point.

Coming back to the ansatz (\ref{Jasymptotic1}), recall
it has  been proven in \cite{Lee:2018urn} that $J_0$ is the class of a holomorphic, rational curve $C_0$ with $C_0 \cdot C_0 = 0$. As such we can parametrise
\be
C_0 = k \, (w_0 + \sum_i c^i \omega_i)
\ee
with $\sum_i (c^i)^2 = 1$ and $k$ some numerical constant. Hence 
\beq
J= t k(\omega_0 + c^i \omega_i) + s_\nu I_\nu \,.
\eeq
We can compare this asymptotic K\"ahler form $J = \sqrt{2 {\cal V}} j$ in the general parametrization (\ref{parametrisation}),
\bea
J &=& \sqrt{2 {\cal V}} \left( \frac{1}{2} e^x (\omega_0 + u^i \omega_i) + \frac{1}{2} e^{-x} (\omega_0 - u^i \omega_i)  \right)   \\
 &=& \sqrt{2 {\cal V}} \left( \frac{1}{2k} e^x J_0+   \frac{1}{2} e^x (u^i - c^i) \omega_i + \frac{1}{2} e^{-x} (\omega_0 - u^i \omega_i)  \right) \label{Jnew2} \,.  
\eea  
Comparison with (\ref{Jasymptotic1}) shows that $e^x \to \infty$ as $t \to \infty$ and we asymptotically identify the prefactor of $J_0$ with the parameter $t$ in (\ref{Jasymptotic1}),
\be
t = \frac{\sqrt{2 {\cal V}}}{2k}   e^x \,.
\ee
The property (\ref{1overtconstraint}) of the limit translates 
 into the requirement  that in particular the second term in (\ref{Jnew2}) must be suppressed by $\frac{1}{t} \sim e^{-x}$ as $x \to \infty$.   
This is possible only if 
\be \label{uiclimit}
u^i(\phi^a) = c^i  + e^{-2x} \tilde u^i(\phi^a) \,,
\ee
where $\tilde u^i(\phi^a)$ takes values at most of order one for $x \to \infty$.   
Note that the condition $\sum_i (c^i)^2 = 1 = \sum_i (u_i)^2$ implies that 
\be
\sum_i c^i \tilde u^i = -\frac{1}{2} e^{-2x} \sum_i \tilde u^i \tilde u^i \,,
\ee
i.e. for $x \to \infty$ the moduli dependent function $\tilde u^i$ must align essentially orthogonally to the parameters $c^i$. This characterizes the limit of nearly vanishing gauge coupling with finite Planck mass, in the chosen field basis.

Importantly, in the limit $x \to \infty$, the canonically normalised scalar $x$ asymptotically controls the gauge kinetic matrix via
\be \label{gYMasymptotic}
f_{ab} = j \cdot C_{ab} =\frac{m_{ab}}{k} e^x + \frac{1}{2} e^{-x} C_{ab} \cdot (\tilde u^i \omega_i+  \omega_0 - u^i \omega_i)   \longrightarrow \frac{m_{ab}}{k} e^x \,.
\ee  
At the same time, in the asymptotic K\"ahler metric
\be \label{C0asym}
{\rm vol}(C_0) = J \cdot C_0 = \sqrt{2 {\cal V}} \,  \left(k  \, e^{-x} + \frac{k}{4} e^{-3x} \sum_i \tilde u^i \tilde u^i (1 - e^{-2x}) \right)
\ee
such that 
\be
{\rm vol}(C_{ab}) \, {\rm vol}(C_0) = (2 {\cal V}) \,  m_{ab} + {\cal O}(e^{-2x}) \,.
\ee

\subsection{Proving the Scalar Weak Gravity Conjecture near the weak coupling point} \label{subsec_proving}

We are  ready to investigate the Weak Gravity Conjecture for several abelian gauge group factors in the weak coupling regime, while systematically including the effect of scalar fields. 
According to the logic of section \ref{sec_SWCG}, the first step consists in identifying the tower of asymptotically massless states near the weak coupling point.
To this end,
recall from the Introduction  that 
a D3-brane wrapping the curve $C_0$ introduced in the previous section gives rise to a critical heterotic string in six dimensions  \cite{Lee:2018urn}. Its tension, in the above conventions, is computed as\footnote{To compare this expression with the conventions used in \cite{Lee:2018urn}, note that we have dropped an overall factor of $2 \pi$ in the underlying ten-dimensional effective action, equ. (A.1) and (A.2) therein.
This corresponds to rescaling the string tension with a factor of $\frac{1}{2\pi}$ compared to the value (A.9) given in \cite{Lee:2018urn}, and furthermore to dropping a factor of $2\pi$ in the mass relation (\ref{massrelation}) compared to equ. (2.52) of  \cite{Lee:2018urn}.}
\be \label{Tformula}
T  = j \cdot C_0  = k e^{-x} + {\cal O}(e^{-3x}) \,,
\ee
where the last equality uses the asymptotic expression (\ref{C0asym}). 
The excitations of this string carry space-time mass
\be \label{massrelation}
M_n^2 = 4 \, T \, (n-1) \,.
\ee
Since this mass vanishes in the weak coupling  limit $x \to \infty$, the associated string excitations represent the sought-after tower of asymptotically massless particles.
In the special case of a single abelian gauge group factor, it was furthermore shown in \cite{Lee:2018urn}
that for each charge in the sublattice defined by $\mathfrak{q}_k  =2 m k$ with $k \in \mathbb Z$, this tower contains a state at excitation level $n(k) = m \, k^2$, i.e. the states in the sublattice satisfy the relation
\be \label{q24mn}
\mathfrak{q}^2_k = 4 \, m \, n(k) \,.
\ee 
The proof of this relation makes use of the arithmetic properties of the elliptic genus and will be reviewed in the next section.

To complete the argument, we must show that the states in the sublattice satisfy the Weak Gravity Conjecture bound. While the final result for the single $U(1)$ case has already been stated in \cite{Lee:2018urn}, a derivation of the effect of scalar fields has not been presented yet, and we provide this derivation now.
However, before coming to this central point of the paper,  we first generalize the above statements to an arbitrary number of abelian gauge group factors: In this case, the matrix $2 \, m_{ab}$ introduced in (\ref{mabdef}) naturally defines a sublattice of the charge lattice $\mathbb Z^{n_V}$,
\be \label{latticemdef}
\Lambda_{2 m} := 2 m \, \mathbb Z^{n_V} := \{\underline{\mathfrak{q}} \in {\mathbb Z}^{n_V} |   \underline{\mathfrak{q}} =  \sum_{a=1}^{n_V} \,  k_a \, \underline{v}_a \, ,\quad k_a \in \mathbb Z  \} \,,
\ee 
where we have expressed the matrix $2 m_{ab}$ as 
\be
2 m = (\underline{v}_1, \ldots, \underline{v}_{n_V}) \,, \qquad \underline{v}_a \in \mathbb Z^{n_V} \,.
\ee
As we will show in the next section, each charge vector $\underline{\mathfrak{q}}(\underline{k})$ in this lattice is populated by a physical state  with associated  excitation level $n(\underline{k})$ such that 
\be \label{qsqaurerel1}
\langle \underline{\mathfrak{q}}, \underline{\mathfrak{q}} \rangle_m :=  m^{ab} \mathfrak{q}_a \mathfrak{q}_b = 4 \, n(\underline{k}) \,.
\ee 
Here $m^{ab}$ is the inverse of the matrix $m_{ab}$. This directly generalizes the expression (\ref{q24mn}) governing the single $U(1)$ case.

In the weak coupling regime $x \to \infty$ we can now trade $m^{ab}$ for $f^{ab}$ with the help of (\ref{gYMasymptotic}). Using also  the mass relation (\ref{massrelation}) along with (\ref{Tformula}) 
shows that the
sublattice of charges $\underline{\mathfrak{q}}(\underline{k})$ is populated by string excitations which obey the charge-to-mass relation
 
\bea
\langle \underline{\mathfrak{q}}, \underline{\mathfrak{q}} \rangle_f = M^2_{n(\underline{k})} + 4m_{ab} f^{ab} >  {M^2_{n(\underline{k})}}  \quad \,{\rm as} \qquad x \to \infty \,.
\eea

If we reinstate the 6d Planck mass, which had been set to unity in the above supergravity analysis, we arrive at
\be \label{chargetomass-Mpl}
\langle \underline{\mathfrak{q}}, \underline{\mathfrak{q}} \rangle_f   >  \frac{M^2_{n(\underline{k})}}{M^4_{\rm Pl}} \quad \,{\rm as} \qquad x \to \infty \,.
\ee
Here $m_{ab} f^{ab} = k e^{-x}$ scales the same way in $x$ as $M_{n(\underline{k})}^2$, but does not grow with $n(\underline{k})$.

Our final goal is to compare the behaviour (\ref{chargetomass-Mpl}) to the WGC bound, including the effect of scalar fields. 
In the weak coupling limit,  the relevant scalar field that controls both $f_{ab}$ and the masses $M^2_{n(\underline{k})}$   of the tower of states in question
 is the canonically normalised field $x$.
The important observation is that, ignoring subleading corrections which vanish as $x \to \infty$, the relevant part of the effective action reduces precisely to that of Einstein-Maxwell-dilaton theory, with dilaton coupling $\alpha=1$,
\be
S= \int_{\mathbb R^{1,5}}      \frac{1}{2}  \sqrt{-g}    R - \frac{1}{2} {\rm d} x \wedge \ast {\rm d} x   -  \frac{1}{2} f_{ab} \, e^{x}  F^a \wedge \ast F^b \,.
\ee
This follows from the general form of the effective action (\ref{pa1})  together with the scalar kinetic matrix (\ref{kinetictermx}) and the gauge kinetic matrix (\ref{gYMasymptotic}), in the limit $x \to \infty$.
The numerical constant (\ref{muGmuphi})  in the extremality bound (\ref{WGCgeneral1}) hence becomes
\be \label{muresult}
\mu = \mu_G + \mu_x =  \frac{3}{4} + \frac{1}{4} = 1 \,,
\ee
in perfect agreement with the observed inequality (\ref{chargetomass-Mpl}). 
The same result follows, of course, from the WGC in its form (\ref{WGC2}) due to the scalar field dependence (\ref{Tformula}) of the masses. 
One can convince oneself that the subleading corrections, in particular those depending on the remaining $n_T-1$ moduli $\phi^a$, do not alter this result in the limit $x \to \infty$.
For our purposes it suffices to determine the scaling of these corrections with $x$.
Explicit computation shows that the contribution  to (\ref{WGC2}) from the extra moduli, identified to first approximation with the order one functions $\tilde u^i$, is suppressed as
\be
\frac{1}{M_{n(\underline{k})}^2} g^{ij} \partial_i M_{n(\underline{k})} \partial_j M_{n(\underline{k})} = {\cal O}(e^{-2x})   \qquad {\rm as} \quad x \to \infty \,.
\ee
Here we made use of the moduli metric (\ref{kinetictermx}).
Hence the result (\ref{muresult}) captures the full scalar field dependence of the WGC bound in the limit $x \to \infty$ under consideration.

As discussed in section \ref{sec_SWCG}, the exponential behaviour of the gauge kinetic matrix near the weak coupling point is guaranteed already on general grounds, if we are willing to accept the Swampland Distance Conjecture as well as the Sublattice WGC.
On the other hand, the precise value  of the dilaton coupling, which governs the exponential behaviour of both the masses of the tower and the gauge kinetic matrix,  can only be determined by an explicit computation in a concrete setup. 
What is remarkable is that this coupling and its resulting value (\ref{muresult}) for the charge-to-mass ratio yields such an accurate lower bound for the physical state excitations in the sublattice (\ref{latticemdef}), near the weak coupling point.
In particular, the underlying relation (\ref{qsqaurerel1}) is purely arithmetic and derives from the modular properties of the elliptic genus. Any other numerical prefactor on the RHS of (\ref{qsqaurerel1}) would have lead to a mismatch with the bound (\ref{muresult}) derived from the Scalar Weak Gravity Conjecture.

Another interesting point to stress is that the contribution $\mu_x = \frac{1}{4}$ to the WGC bound due to the scalar field interactions is of the same order, but slightly smaller than the gravitational contribution $\mu_G = \frac{3}{4}$. Hence at least for the tower of asymptotically massless states considered here, and near $x \to \infty$, the strength of the attractive Yukawa interactions does not exceed that of the gravitational interaction.

\section{\large The Arithmetics of the WGC with multiple U(1) Factors }

It remains to prove the relation (\ref{qsqaurerel1}) used in the previous section, by generalizing our findings of \cite{Lee:2018urn} for the elliptic genus in the presence of a single abelian gauge group to multiple $U(1)$ factors.
After briefly reviewing the role of weak Jacobi forms for a single $U(1)$, we point out the analogous modular properties in the more general situation and finally demonstrate the validity of our general findings for an explicit example.

\subsection{The elliptic genus and higher rank Jacobi forms} \label{subsec_ellgen}

Consider
the critical 6d heterotic string that arises from a D3-brane wrapped on the curve $C_0$ introduced in the previous section. 
If the base $B_2$ is one of the Hirzebruch surfaces, F-theory on $Y_3$ has a perturbative heterotic dual in terms of a standard heterotic compactification on some elliptic  K3-surface, ${\cal K}$. The curve $C_0$ is the fiber of the Hirzebruch surface and the 6d heterotic string in question is \emph{the} heterotic string describing this string vacuum. For bases more general than
Hirzebruch surfaces, 
heterotic duals still exist, but these  are  necessarily strongly coupled as they involve heterotic NS5-branes and therefore produce extra massless, self-dual tensor fields.
We can continue to think of the heterotic duals as compactifications on  K3-surfaces ${\cal K}$, albeit  with extra 5-brane defects. Despite the absence of a fully perturbative description of such theories, quantitative statements are still possible at least for a subsector of the spectrum.

The subsector of the string excitations we have in mind is encoded in the elliptic genus \cite{Schellekens:1986yi,Witten:1986bf}
\be
Z_{\cal K}(\tau, z^a) = {\rm Tr}_R (-1)^F F^2 q^{H_L} \bar q^{H_R} \prod_{a=1}^{n_V} (\xi^a)^{J_a} \,.
\ee
The trace is taken over the Ramond sector of the string along a torus $T^2$, and $F$ denotes the fermion number. The modular parameter $\tau$ of the torus appears in $q = e^{2 \pi i \tau}$.
The $n_V$ abelian gauge factors $U(1)_a$ of the 6d theory act as global symmetries on the worldsheet of the string, with generators $J_a$. The trace is weighted by the $U(1)_a$ charges by including the elliptic parameters $z^a$ in the combination  
\be
\prod_{a=1}^{n_V} (\xi^a)^{J_a} = e^{2\pi i \sum_a J_a z^a } \,,\qquad \xi^a = e^{2 \pi i z^a} \,.
\ee

Using a chain of dualities \cite{Klemm:1996hh,Haghighat:2013gba,Haghighat:2014vxa}, this object can be related \cite{Lee:2018urn} to the generating function for the genus-zero BPS invariants of Gopakumar-Vafa type on the elliptic threefold $Y_3$.
More precisely, the elliptic genus can be computed as 
\be \label{Zkcounitngexp}
Z_{\cal K}(\tau, z^a) =  - q^{-1} {\cal F}_{C_0}^{(0)}(\tau, z^a) =  - q^{-1} \sum N_{C_0}({n,r_a})  q^n \prod_a   (\xi^a)^{r_a} \,.
\ee
Here ${\cal F}_{C_0}^{(0)}(\tau, z^a)$, encoding the BPS invariants $N_{C_0}$, is the genus-zero prepotential of the topological string on the elliptic Calabi-Yau three-fold $Y_3$ as appearing in the expansion of the topological string free energy
\be
{\cal F}(\tau,z^a, t_\beta,\lambda_s) = \sum_{g=0}^\infty \lambda_s^{2g-2} \sum_{C_\beta \in H_2(B_2,\mathbb Z)} {\cal F}^{(g)}_{C_\beta}(\tau, z^a) e^{2\pi i t_\beta} \,.
\ee
The topological string on elliptic three-folds has been studied in detail in the recent literature such as \cite{Klemm:2012sx,Alim:2012ss,Huang:2015sta,DelZotto:2017mee}, to which we refer for more information and references on this vast subject.

The special case of a single abelian group, $n_V=1$, has already been treated in detail in \cite{Lee:2018urn}. In this case, $Z_{\cal K}(\tau, z)$ is a quasi-modular weak Jacobi form (aka {\it quasi-Jacobi form}) of weight $w= - 2$ and fugacity index $m$, where $m = \frac{1}{2} C_0 \cdot C$. 
An analysis of its pole structure \cite{Klemm:2012sx,Alim:2012ss} exhibits that
\be \label{ansatz1}
Z_{\cal K}(\tau, z) =  \frac{ \Phi_{10,m}(\tau,z)    }{\eta^{24}(\tau)}    \,,
\ee
where $\eta(\tau)$ is the Dedekind function and  $\Phi_{10,m}(\tau,z)$ is quasi-modular of weight $w=10$ and fugacity index $m$.
According to a classic result \cite{EichlerZagier} the ring of weak Jacobi forms of modular weight $w$ and fugacity index $m$ is freely-generated by two particular weak Jacobi forms\footnote{See e.g. the appendix of \cite{Lee:2018urn} and references therein for the properties of these functions.}, $\varphi_{0,1}(\tau,z)$ and $\varphi_{-2,1}(\tau,z)$, over the ring of modular forms $\mathbb C[E_4(\tau), E_6(\tau)]$. Here $E_{2k}(\tau)$ denotes the Eisenstein series.
 Quasi-modularity, on the other hand, means, in the present context, that $Z_{\cal K}(\tau, z)$   satisfies the modular anomaly equation  \cite{Klemm:2012sx,Alim:2012ss}
\be \label{holanom}
\frac{\partial}{\partial E_2(\tau)}  {\cal F}_{C_0}^{(0)}(\tau, z) = \frac{1}{24} \sum_{C_1 + C_2 = C_0}   (C_1 \cdot C_2) \,  {\cal F}_{C_1}^{(0)}(\tau, z) {\cal F}_{C_2}^{(0)}(\tau, z) \,.
\ee
The ring of associated functions is generated by the same $\varphi_{0,1}(\tau,z)$ and $\varphi_{-2,1}(\tau,z)$ over the ring $\mathbb C[E_2(\tau), E_4(\tau), E_6(\tau)]$, where $E_2(\tau)$ is only a quasi-modular form. 
In the special case where the base $B_2$ is a Hirzebruch surface and hence a perturbative heterotic dual exists, $Z_{\cal K}(\tau, z)$ is in fact modular, as opposed to quasi-modular, in agreement with classic results  \cite{Schellekens:1986yi} on the elliptic genus of the perturbative heterotic string.

 Given its modular properties, any weak Jacobi form can always
be expanded \cite{EichlerZagier} into a finite number of theta-functions of the form
\be \label{thetaexp1}
\vartheta_{m,\ell}(\tau,z) = \sum_{k\in\IZ}q^{(\ell+2mk)^2/4m}\xi^{\ell+2mk} \,.
\ee
Extremal states are characterized by $\ell=0$ and thus by the vanishing
of the discriminant $\Delta\equiv 4m n_k-\mathfrak{q}_k=0$, which is solved by eq.~(\ref{sublattice1}) and $n_k=mk^2$.
 Due to the mass relation (\ref{massrelation}), this sublattice of physical states indeed has a charge-to-mass ratio that obeys the bound \cite{Lee:2018urn}
\be
g_{\rm YM}^2 \mathfrak{q}_k^2  >  \frac{M^2_{n_k}}{M^4_{\rm Pl}} \quad \,{\rm as} \qquad x \to \infty \,,
\ee
where $\frac{1}{g^2_{\rm YM}}$ replaces $f_{ab}$ in the gauge kinetic term of a single $U(1)$ factor.

After this brief review let us now generalize the above findings to situations with $n_V$ different abelian gauge factors. 
The ansatz (\ref{ansatz1}) is replaced by 
\be
Z_{\cal K}(\tau, z^a) =  \frac{ \Phi_{10,m_{ab}}(\tau, {z}^a)}{\eta^{24}(\tau)}    \,.
\ee
Here $\Phi_{10,m_{ab}}(\tau, {z}^a)$ denotes a lattice quasi-Jacobi form of modular weight $w= 10$
which  is elliptic of rank $n_V$. This means that it depends on $n_V$ elliptic parameters $z^a$. 
The generalisation from rank one to general rank has first been made in \cite{Ziegler} and is  reviewed e.g. in \cite{Oberdieck:2017pqm}.

For simplicity of presentation, we restrict ourselves to modular lattice Jacobi forms for a moment. The modular behavior is now encoded, apart from the modular weight, in the matrix-valued fugacity index $m_{ab}$.
In our context, this matrix is determined geometrically as $m_{ab} = \frac{1}{2} C_{ab} \cdot C_0$.
The transformation properties of a higher-rank modular form $\varphi_{w,m_{ab}}(\tau, {z^a})$ of weight $w$ and fugacity matrix $m_{ab}$ are given by
\bea \label{modformsdef}
\varphi_{w,m_{ab}}\left(\frac{a \tau + b}{ c \tau + d}, \frac{{z}^a}{c\tau +d}\right) &=& (c \tau +d)^w  e^{2\pi i  \frac{c}{c\tau +d}  ({z}^a m_{ab} z^b) }  \varphi_{w,m_{ab}}(\tau, {z}^a) \\
\varphi_{w,m_{ab}}(\tau,z^a + {\lambda}^a \tau + {\mu}^a) &=& e^{-2\pi i  ({\lambda}^a m_{ab} {\lambda}^b \tau +  2 {\lambda}^a  m_{ab} z^b     )}\varphi_{w,m_{ab}}(\tau, {z}^a)
\eea
 where 
 \be
 \lambda^a, \mu^a \in \mathbb Z^{n_V}\,,   \qquad  \left(\begin{array}{cc} a &b \\ c &d \end{array}\right) \in {\rm SL}(2,\mathbb Z) \,.
 \ee
 This definition can be generalized by considering also quasi-modular, lattice Jacobi forms of higher rank, as is relevant in studying F-theory models on a general base.
 The discussion is completely parallel to the rank one case.

As remarked  in the Introduction, higher rank lattice Jacobi forms extend the notion of Weyl invariant Jacobi forms. For the latter, the elliptic variables $z^a$ take values in the complexified Cartan subalgebra of a simple Lie algebra, while the 
elliptic index continues to be determined by a single number \cite{DelZotto:2017mee}. This is a reflection of the Weyl symmetry of Lie algebra weight lattices, but for the more general lattices of the type we consider, there are generically no such symmetries.
Unlike for weak Jacobi forms and some Weyl invariant Jacobi forms, an explicit basis for the space of rank $n_V$ lattice Jacobi forms, comparable to the basis $\varphi_{0,1}(\tau,z)$ and $\varphi_{-2,1}(\tau,z)$ for rank one, is not known to us 
as of this writing, but this is a topic of ongoing research in arithmetic geometry \cite{Zagiertoappear}.
The explicit computation of $Z_{\cal K}(\tau, z^a) $  in concrete examples can therefore be considerably more involved. An example shall be discussed in the next section.

What continues to be available on general grounds, however, is the existence of a finite expansion in terms of lattice theta-functions. This is all we need in order to prove the Sublattice Weak Gravity Conjecture. 
Indeed, every weak Jacobi form with positive definite index matrix $m_{ab}$\footnote{This condition is always satisfied in our context because $m_{ab}$ determines the abelian t'Hooft anomalies on the string, and upon diagonalisation these are positive because they depend quadratically on the charges.} can be expressed 
in terms of 
higher rank theta-functions,  $\vartheta_{m,x_a}(\tau, z^a)$.
 These are defined as
 \be
 \vartheta_{m,x_a}(\tau, z^a)   =  \sum_{\substack{r_a \in \mathbb Z^{n_V} \\ r_a = x_a {\rm mod} \Lambda_{2 m}}}     q^{\frac{1}{4} r_a m^{ab} r_b }   \prod_a (\xi^a)^{r_a}   \,,
 \ee
where the lattice $\Lambda_{2 m}$ has already been introduced in (\ref{latticemdef}).
According to a general theorem (see e.g. the discussion in \cite{Oberdieck:2017pqm}), every quasi-Jacobi form can be written as a sum of finitely many terms, each of the form 
\be
  \left(\xi^1 \frac{d}{d\xi^1}\right)^{d_1} \ldots   \left(\xi^{n_V} \frac{d}{d\xi^{n_V}}\right)^{d_{n_V}}  f_d(\tau, z^a) \,,
\ee
where $d = (d_1,\ldots, d_{n_V})$ and $d_a \in \mathbb Z_{\geq 0}$. Each $f_d(\tau, z^a)$ in turn enjoys an expansion
\be
f_d(\tau,z^a) = \sum_{x_a \in \mathbb Z^{n_V}/\Lambda_{2 m}}   h_{k,x_a}(\tau) \,  \vartheta_{m,x_a}(\tau, z^a) \,.
\ee 
The coefficient functions $h_{k,x_a}(\tau) $ are certain vector-valued modular forms, expressions for which can be found in \cite{Oberdieck:2017pqm} and references therein.

The important point for us is  that the sector with $x_a = 0$ contains states of $U(1)_a$ charges
\be
\mathfrak{q}_a \in \Lambda_{2m}
\ee
 and excitation levels
 \be
 \label{mostextremal}
 n = \frac{1}{4} \mathfrak{q}_a m^{ab} \mathfrak{q}_b \,.
 \ee
This follows by comparing the explicit form of $\vartheta_{m,x_a = 0}(\tau, z^a)$  with the expansion in terms of the state multiplicities as in (\ref{Zkcounitngexp}).
This is exactly the relation (\ref{qsqaurerel1}) needed in order to prove that these states satisfy the Sublattice Weak Gravity Conjecture.

\begin{table} 
\centering
\begin{tabular}{c|c|c|c|c|c|c|c}
 $b_{0}$ & $b_{1}$ & $b_{2}$ & $c_{1}$ & $c_{2}$&$d_{0}$&$d_{1}$&$d_{2}$\\
\hline
 $\alpha-\beta+ \bar {K}$&$\bar {K}$&$-\alpha +\beta+\bar {K}$&$-\alpha+\bar {K}$&$-\beta + \bar {K}$&$\alpha + \bar {K}$&$\beta+\bar {K}$&$\alpha+\beta+\bar {K}$\\
\end{tabular}
 \caption{ Classes of the base sections appearing in the hypersurface equation (\ref{2U1eq1}) (taken from \cite{Borchmann:2013hta}).}\label{coeff}
\end{table}

\subsection{Example: Hirzebruch $\mathbb F_1$ and rank $2$ Mordell-Weil group} \label{subsec_example}

We now exemplify the use of higher rank Jacobi forms in an F-theory model with two abelian gauge group factors. 
This example serves two purposes: First, we will confirm the existence of the sublattice (\ref{latticemdef}) of physical states satisfying the WGC bound;
second, at a technical level, we will see how the intertwining of the two abelian gauge group factors requires a considerably more complicated ansatz to give the elliptic genus $Z_{\cal K}(\tau,z^1,z^2)$ in closed form as compared to a single $U(1)$ factor.
As a main novelty of this example we will present a proposal for a closed expression for the genus-zero Gromov-Witten invariants  and hence the elliptic genus $Z_{\cal K}(\tau,z^1,z^2)$ in terms of a finite set of rank-two lattice Jacobi forms.

A general class of elliptic fibrations which gives rise to two abelian gauge group factors can be expressed as the most generic Calabi-Yau hypersurface in
a ${\rm Bl}^2 \mathbb P^2$ fibration over a general base $B_2$ \cite{Borchmann:2013jwa,Cvetic:2013nia,Cvetic:2013uta,Borchmann:2013hta}.\footnote{This fibration corresponds to the model $F_5$ in \cite{Klevers:2014bqa}. The construction has been generalized further in \cite{Cvetic:2015ioa}.}
In the conventions of \cite{Borchmann:2013jwa,Borchmann:2013hta} we denote the homogenous coordinates of the fiber ambient space by $[u : v : w : s_0 : s_1]$, and the Calabi-Yau hypersurface equation takes the form
\begin{equation} 
\begin{split}
P_{T^2}:= & v\,  w (c_1\,  w\,   s_1 +     c_2\, v\,s_0)  +        u\, (b_0\, v^2\,s_0^2 + b_1\, v\, w\,s_0\,s_1 +  b_2\, w^2\,s_1^2) + \\
&   u^2 (d_0\, v \,s_0^2\,s_1 + d_1\, w \,s_0\,s_1^2 + d_2\, u\,s_0^2\,s_1^2)=0 \,.
\end{split} \label{2U1eq1}
\end{equation}
The objects $c_i$, $b_i$ and $d_i$ transform as sections of certain line bundles on the base $B_2$ which determine the topological structure of the fibration. 
Their transformation behavior is determined by requiring homogeneity of (\ref{2U1eq1}).
A consistent choice for the classes of these base sections has been made in Table \ref{coeff} in terms of the anti-canonical bundle class $\bar K$ of $B_2$ and two more classes $\alpha$ and $\beta$; the latter also determine the transformation behaviour of the  ambient fibral coordinates  as shown in Table \ref{tab-scaling}. They are constrained by the requirement that all classes
appearing in Table \ref{coeff} are non-negative such that holomorphic polynomials $b_i$, $c_i$ and $d_i$ exist.

For generic such base polynomials, the elliptic  fibration $Y_3$ defined by (\ref{2U1eq1})  possesses three independent rational sections. The section divisor associated with  the zero-section can be taken, for instance, to be given by
$S_0: s_0 = 0$, and the divisors associated with two other independent sections as
\be
S_1:   s_1 = 0 \,, \qquad \quad S_2: u=0 \,.
\ee
Then the two independent abelian gauge group factors are generated by the Shioda map\footnote{For background on the construction of abelian gauge symmetries in F-theory, see e.g. the recent reviews \cite{Weigand:2018rez,Cvetic:2018bni}.}
\be
\sigma(S_1) = S_1 - S_0 - \bar K \,, \qquad \quad   \sigma(S_2) = S_2 - S_0 - 2 \bar K + \alpha  \,.
\ee
The most important data for us is the height-pairing matrix
\be \label{Cabinexample}
C_{ab}  = -\pi_\ast (\sigma(S_a) \cdot \sigma(S_b))   =    \left( \begin{array}{cc}   2\bar K &  \bar K + \beta-  \alpha  \\    \bar K + \beta-  \alpha    & 4 \bar K - 2 \alpha \end{array}   \right) \,,
\ee
which determines the gauge kinetic matrix as explained in section \ref{subsec_WGCinFb}. 

Over certain loci in codimension-two on the base $B_2$, the elliptic fiber splits into two fiber components. Apart from the curve class $C_{\mathcal E}$ of the generic fiber, this results in another two independent fibral curve classes $C^{\rm f}_1$ and  $C^{\rm f}_2$. Wrapped M2-branes along the split fiber components over each of these specific loci give rise to matter fields charged under $U(1)_1 \times U(1)_2$. These become massless in the F-theory limit of vanishing fiber volume. For a generic choice of base polynomials, the fiber splitting occurs over six types of codimension-two loci on $B_2$, leading to six species of such charged massless matter \cite{Borchmann:2013jwa,Cvetic:2013nia,Cvetic:2013uta,Borchmann:2013hta},
\bea
&& C_{1,-1} = V(b_0,c_2) \,, \qquad C_{-1,-2} = V(b_2,c_1) \,, \qquad C_{0,2} = V(c_1,c_2) \,,  \label{matter1}\\
&&  C_{1,0} = V_1 \,, \qquad C_{1,1} = V_2 \,, \qquad C_{0,1} = V_3   \label{matter2}  \,.
\eea
Here $C_{\mathfrak{q}_1,\mathfrak{q}_2}$ denotes the locus of matter with charges $\mathfrak{q}_1$ and $\mathfrak{q}_2$, and the notation $V(p_1,p_2)$ refers to the vanishing locus of $p_1$ and $p_2$ on $B_2$, The loci appearing in the second line are associated with more complicated ideals whose form will not be needed.

\begin{table}
\centering
\begin{tabular}{c||ccccc}
 & $u$ & $v$ & $w$ & $s_{0}$ & $s_{1}$\\
\hline
\hline
$\alpha$ & $\cdot$ & $\cdot$ & 1 & $\cdot$ & $\cdot$\\
$\beta$ & $\cdot$ &1 &  $\cdot$ & $\cdot$ & $\cdot$\\
\hline
U & 1 & 1 & 1 & $\cdot$ & $\cdot$ \\
$S_{0}$ & $\cdot$ & $\cdot$ & 1 & 1 & $\cdot$ \\
$S_{1}$ & $\cdot$ & 1 & $\cdot$ & $\cdot$ & 1 \\
\end{tabular}
\caption{Fiber ambient coordinates and their associated divisor classes (taken from \cite{Borchmann:2013hta}).}\label{tab-scaling}
\end{table}

We now specify the base of the fibration to be the
 Hirzebruch surface $B_2 =\mathbb F_a$. A Hirzebruch surface  is a fibration of a $\mathbb P^1$ over $\mathbb P^1$, which can be thought of as the projectivisation of $\cO_{\IP^1} \oplus \cO_{\IP^1}(-a)$. 
The weak coupling limit we aim to analyze corresponds to shrinking the volume of the fiber $\IP^1$, $C_0$, to zero volume while taking the volume of the base $\mathbb P^1$ to be infinite such that the total volume of $B_2 = \mathbb F_a$ stays finite. 
We denote the class of the rational fiber and base of $\mathbb F_a$ class by $f$ and $h$, respectively. Their topological intersection numbers are
\beq
h\cdot h = -a \,,\quad f \cdot f = 0\,,\quad h\cdot f = 1 \,.
\eeq
Additional data of which we will make frequent use are the anti-canonical class of $\mathbb F_a$,
\beq
\bar K= 2h + (2+a) f \,,
\eeq
and the Mori cone $\Mcone$ as well as the closure of the K\"ahler cone $\Kcone$
\bea
\Mcone(\mathbb F_a) &=& {\rm Span}\left<f, h\right> \,, \\
\overline{\Kcone(\mathbb F_a)} &=& {\rm Span}\left<f, h+a f \right> \,.
\eea
To describe the elliptically fibered three-fold $Y_3$, we parametrise the classes $\alpha$ and $\beta$ in Table \ref{coeff} as
\be
\alpha =  (2 - x) h + (2 + a  - y) f \,, \qquad \beta =  (2 - x') h + (2 + a  - y') f \,.
\ee
The base $B_2 = \mathbb F_a$ is itself a toric space with four  toric coordinates $\nu_{z_1}$, \ldots, $\nu_{z_4}$ corresponding to  $\nu_f$, $\nu_{h+af}$, $\nu_f$ and $\nu_h$, respectively. In terms of these and the fiber coordinates,
and modulo taking suitable linear combinations, the GLSM charges of the toric coordinates of the total ambient space of the elliptic fibration can be defined as in Table \ref{tb:F_a_lc-1}.

For the sake of an explicit example, let us specialise to a model over $\mathbb F_{a=1}$ and choose the classes 
\be \label{choicealphabeta}
\alpha =  2h+3f \,,  \qquad \beta=  f\,.
\ee
Note that the class of the polynomial $c_1$ in Table \ref{coeff} becomes trivial. As a result the matter loci associated with the second and third charged matter species in (\ref{matter1}) are absent. 

The resulting height-pairing matrix (\ref{Cabinexample}) implies that the elliptic genus of the string from a D3-brane wrapping $C_0$ transforms as a rank-two weak Jacobi form of fugacity index
\be \label{mabexample}
m_{ab} = \frac{1}{2} C_{ab} \cdot C_0 = \left( \begin{array}{cc}  2 &  0  \\   0 &2  \end{array}   \right) \,.
\ee
The appearance of zeroes on the off-diagonal is an artifact of the choice (\ref{choicealphabeta}), but it does not imply that the two abelian gauge group factors are truly independent because physical states can be simultaneously charged under both factors.

\begin{table}
\begin{center}
\begin{tabular}{c |cccc|ccccc}
& $\nu_{z_1}$ & $\nu_{z_2}$ & $\nu_{z_3}$ & $\nu_{z_4}$ & $\nu_u$ & $\nu_v$ & $\nu_w$ & $\nu_{s_1}$ & $\nu_{s_0}$ \\ \hline
$U(1)_1$ &0&1&0&1&$-2+x$&$x- x' $&0&0&0 \\ \hline
$U(1)_2$ &1&$a$&1&0&$-(2+a)+y $&$ y-y'$&0&0&0\\ \hline
$U(1)_{3}$ &0&0&0&0&1&1&1&0&0\\ \hline
$U(1)_{4}$ &0&0&0&0&$0$&1&$0$&1&0\\ \hline
$U(1)_{5}$ &0&0&0&0&$0$&$0$&1&0&1\\ \hline
\end{tabular}\end{center}
\caption{GLSM charges of the toric coordinates of the ${\rm Bl}^2 \mathbb P^2[3]$ fibration over $\mathbb F_{a}$.} 
\label{tb:F_a_lc-1}
\end{table}

Next let us turn to the computation of the Gromov-Witten invariants $N_{C_0}^{(0)}(n,r_1, r_2)$ of the curve classes
\beq\label{cclass}
\Gamma_{C_0}(n,r_1, r_2)=C_0 + n C_{\cE} + r_1 C^{\rm f}_1 + r_2 C^{\rm f}_2 \,.
\eeq
Here $C_0 = f \subset B_2$ and $C_{\cE}$, $C^{\rm f}_{a=1,2}$ are three fibral curves, chosen such that $C_{\cE}$ represents the full fiber and $S_0 \cdot C^{\rm f}_{a=1,2} = 0$.
The latter property implies that in the M-theory obtained by circle reduction of F-theory on $Y_3$,
an M2-brane wrapping $C^{\rm f}_{a=1,2}$ has vanishing Kaluza-Klein charge.

To compute the BPS invariants via 
 mirror symmetry \cite{Hosono:1993qy,Hosono:1994ax} we need the Mori cone generators as well as the triple intersection numbers, which can be determined via PALP~\cite{Kreuzer:2002uu,Braun:2012vh}. 
We find one toric phase that describes a flat fibration, for which 
the Mori cone and triple intersection numbers take the following form:
\beq\label{mori}
\begin{array}{lclrrrrrrrrrr|}
l^{(1)}&=&(&1,& 0,& 1,& 0,& -1,& 0,& 0,& 0, &0) \\
l^{(2)}&=&(&0,& 1,& -1,& -1,& 0,& 0,& 0,& 1, &0) \\
l^{(3)}&=&(&0,& 0,& -1,& 0,& 1,& 0,& 0,& 0, &1) \\
l^{(4)}&=&(&0,& 0,& 0,& 1,& 0,& 1,& 0,& 0,& -2) \\
l^{(5)}&=&(&0,& 0,& 1,& 0,& 0,& 0,& 1,& 0,& -1) \\
\end{array}
\eeq
\bea\label{inter} \nn
\cI &=&-2 J_1^3-3 J_1^2 J_2+J_1 J_2^2+2 J_2^2 J_3+2 J_2 J_3^2-2 J_1^2 J_4+J_1 J_2 J_4 +2 J_2 J_3 J_4\\ 
&&+\,J_2^2 J_5+J_2 J_3 J_5+J_2 J_4 J_5-3 J_2 J_5^2-2 J_3 J_5^2-2 J_4 J_5^2+8 J_5^3 \,.
\eea
We have characterized the Mori cone generators $l^{(i)}$ by listing their intersection numbers with the $9$ toric divisors $d_\rho = \{{\nu_\rho} = 0\}$, whose  classes are given in Table~\ref{tb:F_a_lc-1}, appropriately reordered. 
These toric divisors can be expressed in terms of $5$ basis elements $J_i$ of $H^{1,1}(X, \IZ)$  as
\bea 
&d_1=-J_1 + 2J_2 + J_3 + J_5 \,,\quad d_2=J_4\,, \quad d_3=-J_1 +J_3 -J_4\,, \quad  
d_4= J_2-J_4\,, \\ \nn  
&d_5=J_1\,,\quad\quad\quad d_6=J_2\,, \quad\quad\quad d_7=J_3\,, \quad\quad\quad d_8=J_4 \,, \quad\quad\quad d_9=J_5 \,. 
\eea

With this input we can compute~\cite{Hosono:1993qy,Hosono:1994ax} the Gromov-Witten invariants of low degree with respect to the basis of curves $l^{(i)}$. The sought-after invariants $N_{C_0}^{(0)}(n,r_1, r_2)$ can then be obtained from them by writing the curve classes $\Gamma_{C_0}(n,r_1, r_2)$ of~\eqref{cclass} in the basis of $l^{(i)}$. 
A detailed analysis shows that 
\beq
C_0 = l^{(4)} \,,\quad
C_\mathcal{E} =2 l^{(1)} + 3 l^{(3)} + 2 l^{(5)} \,,\quad
C^{\rm f}_1 =l^{(1)}+ l^{(3)} + l^{(5)} \,, \quad
C^{\rm f}_2 = l^{(3)} + l^{(5)} \,,
\eeq
and therefore 
\beq
\Gamma_{C_0}(n,r_1,r_2)= (2n+r_1)  l^{(1)} + (3n+r_1+r_2) l^{(3)} + l^{(4)} + (2n+r_1+r_2) l^{(5)} \,.
\eeq
In the end, the relevant Gromov-Witten invariants can be packaged into the following genus-zero prepotential,
\bea
\cF_{C_0}^{(0)}
&\equiv& \sum N_{C_0}^{(0)} (n, r_1, r_2) q^n (\xi^1)^{r_1} (\xi^2)^{r_2} \\ 
&=& -2 +\left[160 + 64 \xi_\pm^{(0,1)} + 64 \xi_\pm^{(1,0)} + 22 \xi_\pm^{(1,1)} + 10 \xi_\pm^{ (1,-1)} \right]q  \label{calFsecondline}\\ \nn
&&+\, \Big[ {52976} + {30400} \xi_\pm^{ (0,1)} + 4928 \xi_\pm^{ (0,2)} + 64 \xi_\pm^{ (0,3)} -2 \xi_\pm^{ (0,4)} + \\ \nn
&& ~~~+\, 30400 \xi_\pm^{(1,0)} + 17028 \xi_\pm^{ (1,1)} + {16956} \xi_\pm^{ (1,-1)} + 2432 \xi_\pm^{(1,2)} + {2432} \xi_\pm^{ (1,-2)} + 10 \xi_\pm^{ (1,3)} +22 \xi_\pm^{ (1,-3)} \\ \nn
&&~~~+\, 4928 \xi_\pm^{(2,0)} + 2432 \xi_\pm^{(2,1)} + 2432 \xi_\pm^{(2,-1)} + 200 \xi_\pm^{(2,2)} + {200} \xi_\pm^{(2,-2)}  \\ \nn
&&~~~+\, 64 \xi_\pm^{(3,0)} + 10 \xi_\pm^{(3,1)} + 22 \xi_\pm^{(3,-1)} - 2 \xi_\pm^{(4,0)} \Big] q^2  \\ \nn 
&&+\, \cO(q^3) \,, 
\eea
with the short-hand notation 
\beq
\xi_\pm^{(r_1, r_2)} := (\xi^1)^{r_1} (\xi^2)^{r_2} + (\xi^1)^{-r_1} (\xi^2)^{-r_2} \,. 
\eeq
The coefficients at level $n=1$ and $(r_1,r_2) \neq (0,0)$ in the first line of (\ref{calFsecondline}) agree with the multiplicities of the charged massless matter fields in the F-theory limit, as derived from (\ref{matter1}) and (\ref{matter2}).
This is because the GW invariants for the split curve components over the codimension-two loci (\ref{matter1}) and (\ref{matter2}) stay the same if we add the full fiber class $C_{\mathcal E}$ and the class $C_0$ in the base. Such pattern is, in fact, required by consistency because the above genus-zero prepotential is related to the elliptic genus of the dual heterotic model as in (\ref{Zkcounitngexp}), and the terms at level $n=1$ give the chiral index of the massless matter content. 
In particular, two of the charge combinations  (\ref{matter1}) and (\ref{matter2}) are absent at this level, as expected based on the discussion after (\ref{choicealphabeta}).

Following the general discussion, we make an ansatz
\beq
\cF_{C_0}^{(0)} (\tau, z^1, z^2) = - \frac{q}{\eta(\tau)^{24}} \Phi_{10, m_{ab}} (\tau, z^1, z^2) \,,
\eeq
with $\Phi_{10, m_{ab}} (\tau, z^1, z^2)$ a rank-two lattice Jacobi form whose index $m_{ab}$ is given in (\ref{mabexample}). 
Note that in the present example $\Phi_{10, m_{ab}} (\tau, z^1, z^2)$ is indeed modular, as opposed to merely quasi-modular, because $C_0$ is the fiber of a Hirzebruch surface and therefore a generator of 
the Mori cone. 
Hence the anomaly equation (\ref{holanom}) of \cite{Klemm:2012sx,Alim:2012ss} is empty and does not imply any dependence on the quasi-modular form, $E_2$.

Unlike for Jacobi forms of rank one, no explicit parametrization of a finite basis for higher rank Jacobi forms seems available in the mathematics literature to date.
The naive assumption would be to continue to work with products of the weak Jacobi forms $\varphi_{-2,1}(\tau, z^a)$ and $\varphi_{0,1}(\tau, z^a)$ for $a=1,2$ (together with $E_4(\tau)$ and $E_6(\tau)$ as always). As it turns out, an ansatz consisting of such a set of functions 
is not sufficient to match the GW invariants computed above. 
What turns out to be successful, on the other hand, is an ansatz for $\Phi$ using  the most general linear combination of the forms\footnote{The forms (\ref{prod}) may not be independent, as explicit analysis suggests.} 
\beq\label{prod}
E_4^{n_4} E_6^{n_6} \prod_{(p,q)} \varphi_{-2,1}(\tau, pz^1+qz^2)^{a_{p,q}} \varphi_{0,1}(\tau, pz^1 + q z^2)^{b_{p,q}} 
\eeq
of weight $10$ and index $m_{ab}$, with $p,q\in \IZ$. Note that it follows from the definition (\ref{modformsdef}) that the index of the product~\eqref{prod} is determined by the quadratic form
\beq
m_{ab} z^a  \, z^b \equiv  a_{p,q} (pz^1 + qz^2)^2  + b_{p,q} (pz^1 + qz^2)^2 \,,
\eeq
which leads to a finite lost of allowed integer pairs $(p,q)$.
Comparison with the Gromov-Witten invariants (\ref{calFsecondline}) with $n=0,1$ already fixes the coefficients in the ansatz for $\Phi \equiv \Phi_{10, m_{ab}} (\tau, z^1, z^2)$ to be  
\bea  \label{fullgen} 
\nn
\Phi &=& 
 \frac{1}{72} E_4^2 E_6 \varphi^{(-)}_{-2,1} \varphi^{(+)}_{-2,1}
+ \frac{1}{1728} E_4^3  \varphi^{(-)}_{-2,1}  \varphi^{(+)}_{0,1}
- \frac{1}{1728} E_6^2 \varphi^{(-)}_{-2,1} \varphi^{(+)}_{-2,1} 
+\frac{1}{62208} E_4^3 E_6 (\varphi^{(1)}_{-2,1})^2 (\varphi^{(2)}_{-2,1})^2 
\\ \nn
&&+\,\frac{5}{62208} E_6^3 (\varphi^{(1)}_{-2,1})^2 (\varphi^{(2)}_{-2,1})^2 
-\frac{1}{27648} E_4^4 (\varphi^{(1)}_{-2,1})^2 \varphi^{(2)}_{-2,1} \varphi^{(2)}_{0,1} 
- \frac{13}{82944} E_4 E_6^2 (\varphi^{(1)}_{-2,1})^2 \varphi^{(1)}_{0,1} (\varphi^{(2)}_{-2,1})^2
\\ \nn
&&
- \, \frac{1}{27648}  E_4^4 (\varphi^{(1)}_{-2,1})^2 \varphi^{(2)}_{-2,1} \varphi^{(2)}_{0,1} 
- \frac{13}{82944} E_4 E_6^2 (\varphi^{(1)}_{-2,1})^2 \varphi^{(2)}_{-2,1} \varphi^{(2)}_{0,1} 
+ \frac{1}{1728} E_4^2 E_6 \varphi^{(1)}_{-2,1} \varphi^{(1)}_{0,1} \varphi^{(2)}_{-2,1} \varphi^{(2)}_{0,1} 
\\ \nn 
&& -\, \frac{29}{248832} E_4^3 (\varphi^{(1)}_{0,1})^2 \varphi^{(2)}_{-2,1} \varphi^{(2)}_{0,1} 
- \frac{19}{248832} E_6^2 (\varphi^{(1)}_{0,1})^2 \varphi^{(2)}_{-2,1} \varphi^{(2)}_{0,1} 
- \frac{29}{248832} E_4^3  \varphi^{(1)}_{-2,1} \varphi^{(1)}_{0,1} (\varphi^{(2)}_{0,1})^2 \\
&&-\, \frac{19}{248832} E_6^2  \varphi^{(1)}_{-2,1} \varphi^{(1)}_{0,1} (\varphi^{(2)}_{0,1})^2
+ \frac{1}{10368} E_4 E_6 (\varphi^{(1)}_{0,1})^2   (\varphi^{(2)}_{0,1})^2 \,,
\eea
where 
\be
\varphi^{(a)}_{*,*} = \varphi_{*,*}(\tau, z^a) \,,  \qquad   \varphi^{(\pm)}_{*,*} = \varphi_{*,*}(\tau, z^1 \pm z^2) \,.
\ee
The analytic expression~\eqref{fullgen} does predict the correct Gromov-Witten invariants~\eqref{calFsecondline} also for $n=2$, as it should. 
We have depicted the lowest-lying charge spectrum encoded in this generating function of Gromov-Witten invariants in Fig.~\ref{f:lattice}.

\begin{figure}[t!]
\centering
\includegraphics[width=14cm] {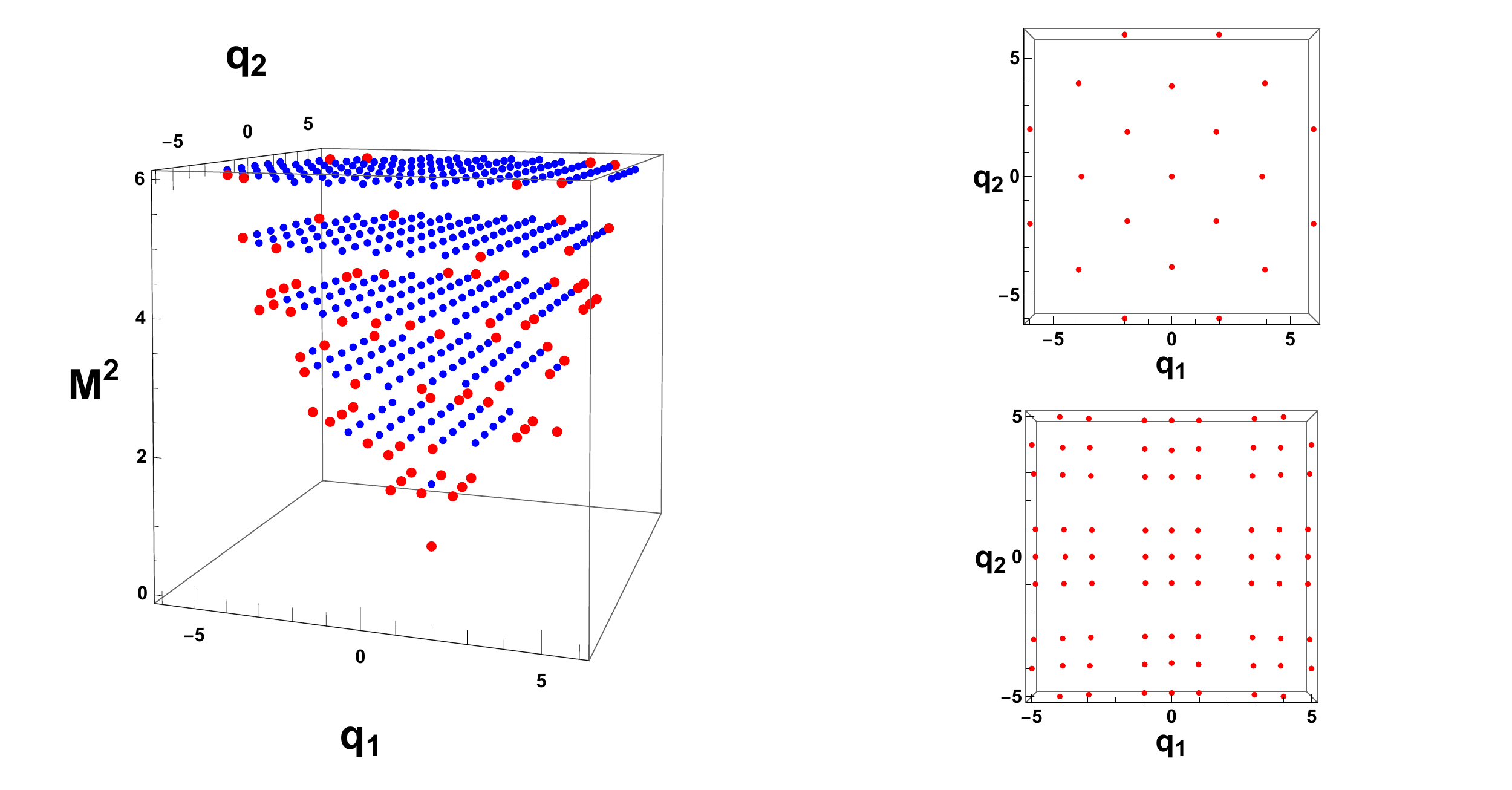}
\caption{ 
Shown on the left is the lowest lying charge-mass spectrum as determined by the elliptic genus given in eq.\ (\ref{fullgen}). The red dots denote
the subset of super-extremal string states.
The top picture on the right shows
the charges of the maximally super-extremal states as determined in (\ref{mostextremal}),  as viewed from the top of the left picture; obviously they lie on a sublattice of the full charge lattice. On the right bottom, the charges of all super-extremal states are shown. These do not form a lattice, rather are given as the union of shifted copies of the sublattice.
}
\label{f:lattice}
\end{figure}

Note that in absence of an underlying mathematical theorem, we cannot strictly guarantee that the ansatz (\ref{prod}) is sufficient and hence that it also captures all higher degree GW invariants beyond the ones computed explicitly via mirror symmetry.
Nonetheless we conjecture that this is the case, motivated by the highly non-trivial checks occurring already at low degree.
Note  again that the non-standard generators $\varphi^{(\pm)}_{*,*}$ were necessary here even though the index $m_{ab}$ is purely diagonal.
It is tempting to speculate that a similar ansatz (\ref{prod}) correctly reproduces the GW invariants also in other situations which require (quasi-)Jacobi forms of higher rank.
This would be an interesting direction for further exploration.

\subsection*{Acknowledgements}

We are grateful to  Vijay Balasubramanian, Mirjam Cveti{\v c}, Roberto Emparan, Gary Gibbons, Steve Giddings, Arthur Hebecker, Nakwoo Kim, Greg Moore, Eran Palti, George Papadopoulos, Harvey Reall and Irene Valenzuela for illuminating discussions and correspondence. 
T.W. thanks the Aspen Center for Physics for hospitality, where parts of this work were carried out.

\bibliography{papers}
\bibliographystyle{JHEP}

\end{document}